\documentclass[a4paper,10pt]{article}
\usepackage{graphicx}% Include figure files
\usepackage{dcolumn}% Align table columns on decimal point
\usepackage{bm}% bold math
\usepackage{amsmath}
\usepackage{amssymb}
\usepackage[mathscr]{eucal}
\begin{document}
\title{Quantum Non-Abelian Hydrodynamics: Anyonic or Spin-Orbital Entangled liquids, Non-Unitarity of Scattering matrix and Charge Fractionalization}
\author{Tribhuvan Prasad Pareek \\
Harish-Chandra Research Institute \\ Chhatnag Road, Jhusi,
Allahabad - 211019, India
}%
\maketitle

\begin{abstract}
In this article we develop an exact(non-adiabatic,non-perturbative) density matrix scattering theory for a two component quantum liquid which interacts or scatters off from a generic spin-dependent quantum potential.
 
The generic spin dependent quantum potential[eq.(1)] is a {\bf matrix potential}, hence, adiabaticity criterion is ill-defined. Therefore the full
{\bf matrix potential} should be treated non-adiabatically. We succeed in doing so using the notion of {\bf vectorial matrices} which allows us to obtain an exact analytical expression for the scattered density matrix ,$\varrho_{sc}$[eq.(\ref{eq:SDM-2})]. We find that the number or charge density in scattered fluid,$\text{Tr}({\varrho}_{sc})$, expressions in eqs. (\ref{eq:DP-1}) depends on {non-trivial quantum interference coefficients}, ${\cal{Q}}^{\alpha\beta}_{0ijk}$ which arises due to quantum interference between spin-independent and
spin-dependent scattering amplitudes and  among spin-dependent scattering amplitudes. 
Further it is shown that $\text{Tr}( {\varrho}_{sc} )$ can be expressed in a compact form [eq.(\ref{eq:DP-1.2})] where the effect of quantum interference coefficients can be include using a  {vector} $\bm{Q}^{\alpha\beta}$ which allows us to define a {\bf vector order parameter} $\bm{Q}$. Since the number density is obtained using and exact scattered density matrix, therefore, we do not need to prove that $\bm{Q}$ is non-zero. However for sake of completeness we make detailed mathematical analysis for the conditions under which the {\bf vector order parameter} $\bm{Q}$ would be zero or non-zero. 

We find that in presence of spin-dependent interaction the {\bf vector order parameter} $\bm{Q}$ is necessarily non-zero and is related to the commutator and anti-commutator of scattering matrix $\cal{S}$ with its dagger $\cal{S}^{\dagger}$[eq.(\ref{Vec-Q})]. It is further shown that $\bm{Q}\neq 0$, implies four physically equivalent conditions,i.e, {\bf spin-orbital entanglement is non-zero}, {\bf Non-Abelian scattering phase ,i.e, matrices}, scattering matrix is {\bf Non-Unitary} and the broken time reversal symmetry for scattered density matrix. This also implies that quasi particle excitation are anyonic in nature, hence, charge fractionalization is a natural consequence. This aspect has also been discussed from the perspective of number or charge density conservation, which implies i.e, $\text{Tr}({\varrho}_{sc})=\text{Tr}({\varrho}_{in})$.
On the other hand $\bm{Q}=0$ turns out to be a mathematically forced unphysical solution in presence of spin-dependent potential or scattering which is equivalent to {\bf Abelian} hydrodynamics ,{\bf Unitary} scattering matrix,  absence of spin-space entanglement, and preserved time reversal symmetry. 

We have formulated the theory using mesoscopic language, specifically, we have considered two terminal systems connected to spin-dependent scattering region, which is equivalent to having two potential wells separated by a generic spin-dependent potential barrier. The formulation using mesoscopic language is practically useful because it leads directly to the measured quantities such as conductance and spin-polarization density in the leads, however, the presented formulation is not limited to the mesoscopic system only, its generality has been stressed at various places in this article.
\end{abstract}

\author{}

\section{Introduction}
Almost Half a century ago in 1957 a seminal paper by Landauer\cite{Landauer} put a theoretical seed for a new direction in condensed matter physics.
However for a theoretical seed to grow  an appropriately fertile experimental land is required. This experimental land was prepared through the painstaking efforts and research of eminent experimentalist and engineers over next 15 years and brought in existence the nano-fabrication or micro-fabrication technology, commonly know as MBE, for a beautiful account of such efforts see Ref.\cite{Patrick-MBE}. Availability of fertile land made the seed to sprout and in its early days the name, 'Mesoscopic' was give by Van Kampen\cite{Van-Kampen-Book}. This early sprout was further watered by B\"uttiker through another important theoretical paper which cleared the role of reservoir particularly as a source of dephasing\cite{Buttiker}. This lead immense growth of the field which today is known as mesoscopic physic\cite{Van-Kampen-Book,Datta-Book,Imery-Book,Ferry-Book}. Theoretically and experimentally entangled advances in mesoscopic physics during the next 
three decades revealed Quantum Coherent Charge Transport(QCCT) phenomena such as, one-dimensional localization or series addition of external-quantum resistors\cite{Landauer-series,Anderson-series}, Aharonov-Bohm(AB) Oscillation in ring shaped conductors or parallel addition of external-quantum resistors \cite{Gefen-parallel,Webb-AB},
Universal conductance fluctuation\cite{Altshuler-UCF,Stone-UCF} etc. to name but a few. All these experimental phenomena are quantum interference effects due to spatial(orbital) part of electronic wave-function where 
the spin part of wave function or spin (internal) degree of freedom did not play an active role since scattering is
spin independent. Therefore electronic phase coherence is solely determined by the spatial(orbital) phase coherence length $L_{\phi}$ which 
is temperature dependent because only inelastic scattering could cause random phase changes in an irreversible which
leads to decoherence. On the other hand elastic scattering
changes the phase in a deterministic and reversible way therefore does not causes decoherence. Hence a mesoscopic sample of effective physical dimension of ($L$) is phase coherent as long as $L \leq L_{\phi}$.

The presence of spin dependent elastic scattering (spin conserving and spin-flip scatterings) causes spin to play an active role.  
Under generic conditions spin dependent elastic scattering can occur due to presence of spin-orbit interaction, static magnetic impurities, non-collinearity of magnetization direction in mesoscopic multilayer systems and inhomogeneous magnetization profile in bulk(macroscopic system)\cite{Baibich-GMR,Moodera-TMR,Mcguire-AMR, Dieny-SVE,Silsbee-inj,Schmidt-inj,Rashba-inj,Fert-inj,Smith-inj,Jedema-inj,Lou-inj}.
Theoretically spin-dependent transport in presence of spin-dependent elastic scattering has been studied using various approaches such as, drift-diffusion model of spin transport\cite{Zutic-sd,Fabian-sd,DasSarma-sd,Halperin-sd},Boltzmann approach\cite{Sinitsyn-BZ,Culcer-BZ,Niu-BZ}, Kubo approach\cite{Sinova-SHE-KB,Tse-KB,Adeline-KB} and Landauer-B\"{u}ttiker formalism \cite{Brataas-LB,pareek-LB,Molenkamp-LB,Aharony-LB,Bardarson-LB,Ren-LB,Nikolic-LB,Loss-LB,scheid-LB}. 
All these studies treats the spin-dependent quantum scattering potential in an adiabatic approximation, which amounts to two steps, namely, (i) choosing a specific spin quantization axis (ii) replacing the spin dependent quantum potential along the chosen axis by its average classical value. To understand this more clearly consider the 
generic quantum scattering potential for electrons which can always be decomposed in electronic spin space as(see section 3 in main text),
\begin{equation} 
V(r)= V_{0}(r) {I}_{2}^{s} + \bm{V}(r) \cdot \bm{\sigma} 
\label{eq:int-1}
\end{equation}
where $V_{0}$ is spin-independent scattering potential, $\bm{V}=V_{x}\hat{x}+V_{y}\hat{y}+V_{z}\hat{z}$ is spin dependent scattering potential, ${I}_{2}^{s}$ is $2\times 2$ identity matrix in spin space and $\bm{\sigma}$ is vector of Pauli matrices. 
To make an adiabatic approximation for this potential let us choose spin quantization axis along ${\hat{\bm{z}}}$ axis and replace $V_{z}$ by its average classical value,i.e., $V_{z}\approx \langle V_{z} \rangle $, this leads to
\begin{equation}
 V(r)\approx V_{adia}(r)= V_{0}(r) {I}_{2}^{s} + V_{x}\sigma_{x}+V_{y}\sigma_{y}+ \langle V_{z} \rangle \sigma_{x}.
\label{eq:int-2}
\end{equation}
In the above $ \langle V_{z} \rangle$ acts as a classical magnetic field. In other words under the {\bf adiabatic approximation} a part of the
spin-dependent quantum scattering potential acts as an effective classical magnetic field
($B_{int}=\langle V_{z} \rangle$) in the reference frame of the moving electrons which in-turn allows to treat the electron gas residing in the scattering media as a two component classical fluid, for a clear and concise discussion of this see Ref.\cite{Stern-AB-SO}.
Experimental validity of this adiabatic perturbative approach is best seen in the beating pattern observed in the magneto-conductance oscillation of mesoscopic ring in presence of 
spin-orbit coupling and external magnetic filed\cite{Yau-AB-SO,Boris-AB-SO,Bergsten-AC-SO}. These beating pattern arises due to superposition of 
two frequencies at $\phi_{AB}\pm \phi_{SO}$, where  $\phi_{AB}$ is Aharonov-Bohm Phase and $ \phi_{SO}$ is spin geometric phase. It is important to note that the AB phase $\phi_{AB}$ is same for the two components while the spin geometric phase is $+\phi_{SO}$ for the spin up component and $-\phi_{SO}$ for the spin down component in accordance with the classical binary nature of spin. This implies that within this adiabatic/classical approximation the spin-geometric phase becomes an abelian quantity because $\phi_{SO}$ turns out to be a number and not a matrix, while the full quantum scattering potential given in eq.(1) is a $2\times 2$ {\bf matrix potential} ,therefore, the scattering phase associated with it should also be a matrix which would make it a {\bf non-abelian phase}. In fact the {\bf non-abelian}  nature of scattering phases has been 
at the root of theoretical problem of defining spin currents and its conservation laws both quantum mechanically \cite{Bader-scur,Sinova-scur,Jungwirth-scur,Rashba-scur,Shi-scur,Shen-scur,Wang-scur,Chen-scur,Sonin-scur} as well within the gauge theoretical formulation \cite{Frohlich,Leurs,Tokatly}.

Before we proceed let us re-look at {\bf adiabatic criterion} semi-quantitatively. If the potential is spin-independent, i.e, $\bm{V}=0$ in eq.(\ref{eq:int-1}) then one can think of switching on the spin-independent potential $V_{0}(r)$ slowly. This adiabatic switching on of the interaction is formulated mathematically by including a factor $exp(-\epsilon_{0} |t|)$ in the scalar potential $V_{0}(r)$ and allow $\epsilon_{0} \mapsto 0$ at the end of calculation. This works as long as potential is spin-independent. However as soon as spin-dependent potential is included the potential is no longer a scalar function rather it becomes a $2 \times 2$ matrix see eq.(\ref{eq:int-1}), consequently,
adiabatic switching on of the full matrix potential becomes ill defined. This is so because matrix potential has four elements which can have independent time scales about which we do not have any knowledge a priory. However a perturbative adiabatic progress can still be made by choosing
a spin-quantization axis(say along $z$ axis) which is physically equivalent to applying a tiny external magnetic field in an adiabatic way,i.e, $B_{ext,z}\equiv B_{ext,z}exp(-\epsilon_{z} |t|)$ [$B_{ext,z} \ll \langle V_{z} \rangle$]. This amounts to replacing the $V_{z}$ by
$V_{z}\mapsto \langle V_{z} \rangle + B_{ext,z}\approx \langle V_{z} \rangle$ as is done in eq.(\ref{eq:int-2}). 
Further if one assumes $\epsilon_{0}=\epsilon_{z}$ that would imply that both spin-independent and spin-dependent potential along the chosen quantization axis can be switched on adiabatically simultaneously. Therefore within this approximation the time scales associated with the off-diagonal elements, i.e., $V_{x}$ and $V_{y}$ only gives rise to small quantum fluctuations. However in reality one does not have a prior knowledge of the time scales associated with the spin-independent scalar potential $V_{0}$ and spin dependent vector potential $\bm{V}$, which is equivalent to having a $ 2 \times 2 $ matrix potential. Therefore a {\bf non-adiabatic} treatment of the full quantum potential requires that one should not chooses a spin-quantization axis from the very beginning and treat the potential as a {\bf matrix potential}.

From the discussion above we see that a non-adiabatic, non-abelian treatment of spin-dependent scattering requires that one should take into account the quantum scattering potential given in eq.(1) instead of the approximate potential  given in eq.(2).
This means that one should use a formulation which allows to treat all spin-quantization axis at equal footing which is possible only in density matrix formulation. Therefore in this article we develop an exact analytical density matrix scattering theory of spin dependent scattering in hydrodynamic(long wavelength) regime. In particular starting from a two component quantum liquid described by an incident density matrix defined in section \ref{sub-sec-R} and \ref{sub-sec-R-IDM} main text, we include the effect of matrix potential(discussed \ref{sec-SPSM}) using scattering matrix,${\cal{S}}$, defined in subsection (\ref{sec-SPSM}) in a non-adiabatic, non perturbative way and calculate the scattered density matrix exactly in hydrodynamic regime using the relation $\rho_{sc}={\cal{S}}\rho_{in}{\cal{S}}^{\dagger}$,the expressions(\ref{eq:SDM-1}-\ref{eq:SDM-3}) in section \ref{sec:SDM}. 
In section(\ref{sec-Num-DP}) the number or charge density in scattered fluid is obtained which is given by $\text{Tr}(\rho_{sc})$ [equivalent expressions (\ref{eq:DP-1},\ref{eq:DP-1.1},\ref{eq:DP-1.2})]. Similarly the spin-density vector for scattered fluid is given by $\text{Tr}(\rho_{sc}\bm{\sigma})\equiv \{\text{Tr}(\rho_{sc}\sigma_{x}),\text{Tr}(\rho_{sc}\sigma_{y}),\text{Tr}(\rho_{sc}\sigma_{z})\}$ in
eqs.(\ref{eq:DP-SD},\ref{eq:DP-SD-1}). 

We find that both number(charge) density as well the spin-density in scattered fluid has contribution which arise due to {\it quantum interference}
between spin-independent and spin-dependent scattering matrix elements. In particular the number density expression (\ref{eq:DP-1}) has {\it quantum interference coefficients} ${\cal{Q}}^{\alpha\beta}_{0ijk}$,\{beside the conventional unitary coefficients 
${\cal{C}}^{\alpha\beta}=(|S_{0,\alpha \beta}|^{2} +\sum_{i}|S_{i,\alpha \beta}|^{2})$ which are sum of scattering probabilities \} which are given by,
\begin{equation}
{\cal{Q}}_{0ijk}^{\alpha\beta}=[\text{Re}(S_{0,\alpha \beta}S_{i,\alpha \beta}^{*})-
\text{Im}(S_{j,\alpha\beta}^{*} S_{k,\alpha\beta})]
\end{equation}
where $S_{0,\alpha \beta}$ is spin-independent scattering amplitudes and $[S_{i,\alpha \beta},S_{j,\alpha \beta},S_{k,\alpha \beta}]$ are spin-dependent scattering amplitudes. Here $[(\alpha,\beta) \in (L, R)]$ are lead index denoting left and right lead (equivalently spin-independent potential wells see Fig.(2)),and  index ${(i,j,k)}$ denote three spin-components in Laboratory Frame.
{\it  Importantly we note that since these quantum coefficients appears in $\text{Tr}(\rho_{sc})$, where $(\rho_{sc})$ has been calculated exactly, therefore, these quantum coefficients are necessarily non-zero and arise due to {\bf spin-space entanglement}}. However for sake of clarity and completeness  we make a detailed analysis of various conditions under which {\it quantum coefficient} ${\cal{Q}}^{\alpha\beta}_{0ijk}$ are zero or non zero in section (\ref{AB-NAB}). We find that taking ${\cal{Q}}^{\alpha\beta}_{0ijk}=0$ requires that scattering matrix elements be {\bf real quaternion}\cite{Dyson-Mehta,Altland,RM-Beena}.\footnote[1]{A complex quaternion is defined as, $q=q_{0}\bm{I}_{2}+\bm{q}\cdot \bm{\sigma}$, where $q_{0}$ is a complex number and $\bm{q}=\sum_{i} q_{i}\hat{i}$ is a complex vector,i.e, $q_{i}$'s are complex number.  A real quaternion corresponds to choosing $q_{i}$' to be pure imaginary,i.e, $q_{i}\equiv \bm{\dot{\iota}}q_{i}$, with $q_{i}$'s now real.}.
A real quaternionic scattering element is composed of four real number corresponding to four scattering amplitudes, one spin-independent and three spin-dependent scattering amplitudes, and a {\bf fixed relative phase between} spin-independent and spin-dependent scattering amplitude(imaginary unit $\bm{\dot{\iota}}$ in eq.(\ref{quaternion}), which is equivalent to having a single effective scattering phase and hence is equivalent to {\bf Abelian hydrodynamics}. The {\bf real quaternionic scattering matrix}  has been extensively used in past for studying symmetry classification of random matrices\cite{Dyson-Mehta, Altland} and it application to charge transport phenomena in mesoscopic systems\cite{RM-Beena} and has been also used to study spin-transport phenomena during past decade\cite{Kirch}.
On the other hand ${\cal{Q}}^{\alpha\beta}_{0ijk}\neq 0$ implies that ``scattering phase'' are complex quaternion which are complex $2 \times 2$ matrices with four independent complex numbers, therefore scattering phases are $2 \times 2$ matrices hence corresponds to {\bf Non-Abelian hydrodynamics}.
 
Moreover the rigorous mathematical analysis presented in sec.(\ref{AB-NAB}) also shows that ${\cal{Q}}^{\alpha\beta}_{0ijk}\neq 0$ also implies that scattering matrix is {\bf non-unitary}. Further it is shown that effect of {\it quantum interference coefficients} ${\cal{Q}}^{\alpha\beta}_{0ijk}$,  can be take into account by defining a {\bf Vector Order Parmeter} $\bm{Q}$ which is related to the scattering matrix in the following way (The eq.(\ref{Vec-Q}) in sec.(\ref{AB-NAB})) ,
\begin{eqnarray}
2\bm{Q}&=& \text{Tr}[({\cal{S}}{\cal{S}}^{\dagger}+{\cal{S}}^{\dagger}{\cal{S}})\bm{\sigma}]+\text{Tr}[({\cal{S}}^{\dagger}{\cal{S}}-{\cal{S}}{\cal{S}}^{\dagger})\bm{\sigma}]\nonumber \\
&=&\text{Tr}\left[[{\cal{S}}^{\dagger},{\cal{S}}]_{+} \bm{\sigma}\right]+\text{Tr}\left[[{\cal{S}}^{\dagger},{\cal{S}}]_{-} \bm{\sigma}\right],
\end{eqnarray}
A non-zero value of  $\bm{Q}$ defines the three physical equivalent notions, namely, non-unitarity of scattering matrix, non-abelian hydrodynamics and non-zero spin-orbital entanglement. An illustration of this is provided in Fig.(4).

The non-unitarity of scattering matrix can also be seen from a simple observation, though it is not rigorously correct. {\bf Toward this end we note that if we take $\rho_{in}$ to be an identity matrix then the scattered density matrix is nothing but the hermitian matrix $\rho_{sc}={\cal{S}}{\cal{S}}^{\dagger}$, hinting that scattering matrix neither needs to be unitary nor hermitian, rather the only requirement is that the matrix ${\cal{S}}{\cal{S}}^{\dagger}$ should be hermitian}. In fact demanding unitarity of scattering matrix amounts to neglecting off-diagonal elements of scattered density matrix which are related to the {\it quantum coefficient} ${\cal{Q}}^{\alpha\beta}_{0ijk}$ which appears in the charge density, $\text{Tr}(\rho_{sc})$, of scattered fluid eqs.(\ref{eq:DP-1}-\ref{eq:DP-1.1}).

In section(\ref{GLCL}) we look at the above discussed various aspect from the point of vies of conservation laws: number(charge) density conservation and show that {\bf charge fractionalization} is a natural consequence. In section (\ref{sec:time-rev}) time reversal operation is defined and shown that scattered density matrix explicitly breaks the time reversal symmetry. 

Finally before we start our main journey let us make a list of essential mathematical tools which would be required to perform the analytical calculation in a compact and transparent way. To make the journey mathematically less tiring we only use vector and matrix algebra separately and in hybridized form which leads to new mathematical objects namely, {\bf vectorial matrices}(introduced in following subsection) in spatial and spin space beside the conventional mathematical objects such as scalar matrices, simple vectors and scalars. Therefore we briefly define and introduce notational convention for these {\bf vectorial matrices} and various mathematical operations on them and among them in the subsection below. 
Throughout the main article most often we quote the final results and give the calculations details in the appendix (\ref{appendix1}).

\subsection{Notational Convention and Vectorial Matrices:} 
\label{vec-M}
The bold faced symbol(Greek letter iota) $\bm{\dot{\iota}}$ would be reserved for the imaginary unit, i.e., $\bm{\dot{\iota}}=\sqrt{-1}$.
The Greek symbols ($\alpha,\beta,\gamma..)$ would be used to denote the leads while lower case alphabets $(i,j,k...)$ will be used to denote the Cartesian axis of laboratory frame in which measurement is performed(see Fig.(1)). Unit vector along Cartesian direction $i$ would be denoted by the symbol $\hat{i}$. The symbol $\{\varepsilon_{ijk},(i,j,k) \in (x,y,z)\}$ would denote completely antisymmetric Levi-Civita symbol in three dimensions.  In summation symbols whenever $i,j,k$ occurs without any braces around them, i.e., $\sum_{ijk}$ , this would imply sum over all permutations(cyclic and anti-cyclic) of $(x,y,z)$, i.e.,  $\sum_{ijk} \equiv [(x,y,z),(y,z,x),(z,x,y),(x,z,y),(z,y,x),(y,x,z)]$. On the other hand a small bracket with subscript 'c' around $i,j,k$, i.e, the symbol  $(i,j,k)_{c}$ would imply summation over cyclic permutations only in other words
$\sum_{(ijk)_{c}} \in [(x,y,z),(y,z,x),(z,x,y)]$. 
The trace symbol $\text{Tr}$ would imply trace over, Cartesian, Leads and the spin space.

The commutator and anti-commutator would be denote by the symbols, $[- , -]_{-}$ and $[- , -]_{+}$ respectively.

A {\bf spatial vectorial matrices},i.e., vectorial matrices in spatial space, would be denoted by upper case boldface letters such as
$ \{ \bm{A},\bm{B},.. \}$ etc. More precisely the vectorial matrix, $\bm{A}$, is defined as $\bm{A}=\sum_{i}A_{i}\hat{i}\equiv A_{x}\hat{x}+A_{y}\hat{y}+A_{z}\hat{z}$, where $A_{i}$ are square scalar matrices of dimension $N$. The elements of matrices $A_{i}$'s are $a_{i, lm} \in   A_{i}$ with  $a_{i, lm}$ are complex or real scalars or functions. Throughout this article we will be considering only square matrices of dimension $N$. The identity matrix in spatial space would be denoted as $I$. 

%Therefore the {\bf vectorial identity matrices} in spatial space corresponds to $\bm{I}=I [\sum_{i}\hat{i}]\equiv I [\hat{x}+\hat{y}+\hat{z}]$. For sake of completeness we mention that the trace of a spatial vectorial matrix is a vector(complex or real) while that of scalar matrix is a number(complex or real). Similarly the trace of {\bf vectorial identity matrices} $\bm{I}$ of dimension $N$ is vector $N\sum_{i}$.

Similarly in the spin space we have four scalar matrices,i.e, $\{I_{2}^{s}, \sigma_{x},\sigma_{y},\sigma_{z}\}$, where $I_{2}^{s}$ and $\{\sigma_{i}, i \in (x,y,z)\}$ are $2 \times 2$ identity and Pauli matrices in spin space respectively. 
The Pauli vector defined as, $\bm{\sigma}=\sigma_{x}\hat{x}+\sigma_{y}\hat{y}+\sigma_{z}\hat{z}\equiv \sum_{i}\sigma_{i}\hat{i}$, is nothing but a
$2\times 2$ vectorial matrix in spin space or {\bf spin vectorial matrix} in our language.  
%The formally one can also define two dimensional {\bf vectorial identity matrices} in spin space as $\bm{I}_{2}^{s}= I_{2}^{s}[\sum_{i}\hat{i}]\equiv I_{2}^{s}[\hat{x}+\hat{y}+\hat{z}]$. 

Now we define mathematical operations such as dot, cross and scalar triple products of {\bf spatial vectorial matrices}. First let us consider three spatial vectorial matrices $\{ \bm{A},\bm{B},\bm{C} \}$, the dot, cross and scalar triple product among these matrices is defined as,
\begin{eqnarray}
 \bm{A}\cdot\bm{B}=\sum_{i}A_{i}B_{i}\neq \sum_{l}B_{i}A_{i}=\bm{B}\cdot\bm{A},
\label{eq:NC-1} \\
\bm{A}\times\bm{B}=\sum_{ijk}\varepsilon_{ijk}A_{j}B_{k}\hat{i} \neq -\bm{B}\times\bm{A}=-\sum_{ijk}\varepsilon_{ijk}B_{j}A_{k}\hat{i},
\label{eq:NC-2} \\
\bm{A}\times\bm{A}=\sum_{ijk}\varepsilon_{ijk}A_{j}A_{k}\hat{i}\neq 0 \label{eq:NC-3}\\
 (\bm{A}\times \bm{B})\cdot \bm{C}=\sum_{ijk}\varepsilon_{ijk}A_{j}B_{k}C_{i}=\sum_{ijk}\varepsilon_{ijk}A_{j}B_{k}C_{i} \label{eq:NC-4}.
\end{eqnarray}
From expression (\ref{eq:NC-1}) we see that the dot product of two spatial vectorial matrices is non-commutative due to
non-commutativity of matrix multiplication of scalar matrices. The non-commutativity of matrix product also shows up in cross product eq.(\ref{eq:NC-2}) above where,$\bm{A}\times\bm{B}$ in general has no simple relation with $\bm{B}\times\bm{A}$ and $\bm{A}\times\bm{A}\neq 0$ in contrast to simple vectors accordingly the cross product of a vectorial matrix $\bm{A}$ with itself is in general need not to be zero as is seen in expression(\ref{eq:NC-3}) above. These non-commutative properties play an important role in throughout this article. For sake of clarity we express the cross product of two vectorial matrices in a more useful form as follows,
\begin{eqnarray}
 \bm{A}\times\bm{B}=\sum_{ijk}\varepsilon_{ijk}A_{j}B_{k}\hat{i}=\sum_{(ijk)_{c}}(A_{j}B_{k}-A_{k}B_{j}) \hat{i} \label{eq:NC-2-1}
\end{eqnarray}
In the rest of the text we use this form of cross product quite often to simply the complicated expressions.

Finally the dot product of a spatial square vectorial matrix $\bm{A}$ of dimension $N$ with Pauli vectorial matrix  $\bm{\sigma}$ is defined via direct product of constituent scalar matrices as ,
\begin{equation}
 \bm{A}\cdot\bm{\sigma}=\sum_{i}A_{i}\sigma_{i}\equiv\sum_{i\in\{x,y,z\}}A_{i}\otimes\sigma_{i}= \bm{\sigma}\cdot \bm{A},
\label{eq:NC-5}
\end{equation}
where $A_{i}\otimes\sigma_{i}$ is a scalar matrix of dimension $2N$.
We note that since  $\sigma_{i}$ and $A_{i}$ acts on
different subspaces, namely,  spin space(internal degrees) and spatial space(orbital degrees) respectively and have different dimensions,therefore, the product, $A_{i}\sigma_{i}$ is defined only in terms of the direct product, i.e., $A_{i}\sigma_{i}\equiv A_{i}\otimes\sigma_{i}$,hence, $A_{i}\sigma_{i} $ is $2N$ dimensional matrix for each $i$. 
Moreover because of the same reasons this direct product is commutative, i.e., $\sigma_{i}A_{i}\equiv \sigma_{i}\otimes A_{i}=A_{i}\otimes \sigma_{i} $. Therefore the dot product between a 
{\bf spatial vectorial matrix} and {\bf Pauli Vector} gives rise to scalar matrix. The defining eq.(\ref{eq:NC-5})  and the following identities
are important which are used quite frequently in the rest of the article. 
\begin{eqnarray}
(\bm{\sigma}\cdot \bm{A})(\bm{\sigma}\cdot \bm{B})=(\bm{A}\cdot\bm{B})\bm{I}^{s}+i(\bm{A}\times\bm{B})\cdot \bm{\sigma} \\
 ( \bm{\sigma}\cdot \bm{B}) ( \bm{\sigma}\cdot \bm{A})=(\bm{B}\cdot\bm{A})+i\bm{\sigma}\cdot (\bm{B}\times \bm{A}) \neq 
( \bm{\sigma}\cdot \bm{A}) ( \bm{\sigma}\cdot \bm{B}), \label{eq:NC-a} \\
( \bm{\sigma}\cdot \bm{A}) ( \bm{\sigma}\cdot \bm{A})=(\bm{A}\cdot\bm{A})+i\bm{\sigma}\cdot (\bm{A}\times \bm{A}) \label{eq:Nc-b}
\end{eqnarray}
We notice that since $\bm{A}$ and $\bm{B}$ are vectorial matrices hence we have the relation (\ref{eq:NC-a}) which markedly different from the text book identities for simple vectors and arises due to {\bf Non-Commutativity of scalar and cross product of vectorial matrices} which plays an important role for non-abelian hydrodynamics. For sake of completeness we have provided the details in appendix.

Other quantities or expression which appears beside these in the rest of this article there meaning would either be clear by the context otherwise it would be stated explicitly.

\section{The Reservoirs and Incident Density Matrix}
\label{sec-R}
A typical two terminal mesoscopic systems consists of two parts namely, the electronic reservoirs and their adiabatic extension know as ideal leads and the mesoscopic sample,i.e, the elastic scattering region shown as a circle in Fig.~(2a). 
The fixed {\bf laboratory frame}  in which measurement is done is shown in the left upper most panel of Fig.~(2b). In this fixed laboratory frame the ideal leads are extended along $\hat{x}$ axis, i.e, this is the direction of particle current flow(longitudinal direction) while along the transverses direction the system is confined. This system is equivalent to two spin-independent potential wells separated by a spin-dependent potential well, this equivalence will be discussed further in sec.(\ref{sub-sec-R-IDM}).
In the following we describe the states of various parts of mesoscopic system in the laboratory frame.

\begin{center}
\begin{figure}
\includegraphics[width=\linewidth,height=3.0in,angle=0,keepaspectratio]{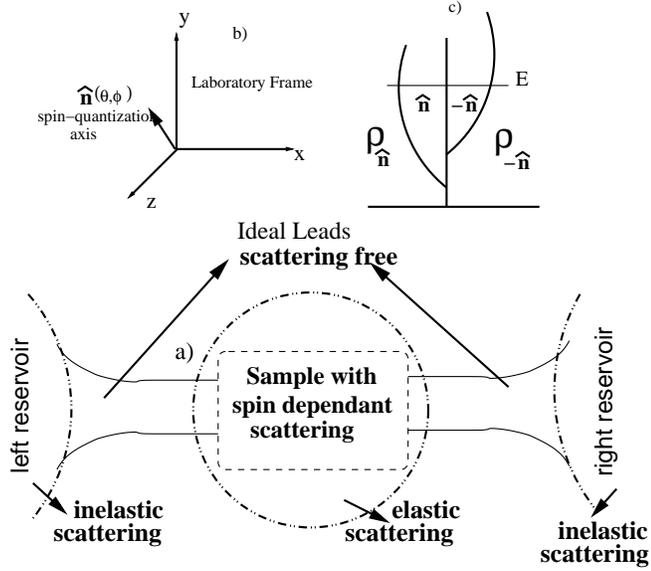}
\caption{\label{fig3} Schematic Figure (a) The two terminal mesoscopic system with spatial separation of scattering regions as shown above and explained in the text. (b) the Fixed laboratory frame, in this frame leads are parallel to $\hat{x}$ axis. (c) the up and down density of two component electronic liquid.}
\end{figure}
\end{center}

\subsection{The Reservoirs: Magnetic and Non-Magnetic}
\label{sub-sec-R}
To describe magnetic and non-magnetic reservoirs at equal footing 
we consider the reservoirs as a source of two component quantum electronic fluid corresponding to spin degree of freedom(see Fig.~(2c)). 
A magnetic reservoir with magnetization direction along an axis  $\bm{\hat{n}}$ defined by the polar and azimuthal angle $\theta$ and $\phi$ in the laboratory frame, can be described as composed  of two densities, $\rho_{\bm{\hat{n}}}$, and $\rho_{-\bm{\hat{n}}}$, corresponding to up and down spin densities of the electronic fluid, where . Therefore the total density and the net spin-density(equivalently net spin polarization or magnetization) of electronic fluid is given by, $\rho_{0}=\rho_{\bm{\hat{n}}}+\rho_{\bm{-\hat{n}}}$ and $\bm{p}=p\bm{\hat{n}}=\rho_{\bm{\hat{n}}}-\rho_{\bm{-\hat{n}}}\equiv P(\cos(\phi)\sin(\theta),\sin(\phi)\sin(\theta),\cos(\theta))$, with $P$ being magnitude of polarization(magnetization) which is uniform(independent of polar angle $\theta$ and azimuthal angle $\phi$). The corresponding density matrix at a fixed total energy $E$ is,
\begin{equation}
\varrho(E,\hat{\bm{n}})=(1/2) (\rho_{0}{I}_{2}^{s}+\bm{p}\cdot\bm{\sigma}).
\label{eq:Res-1}
\end{equation}

The number density and polarization vector corresponding to above density matrix is related to the traces, $\text{Tr}(\varrho(E,\hat{\bm{n}}))=\rho_{0}$ and $\text{Tr}(\rho(E,\hat{\bm{n}})\bm{\sigma})=\bm{p}$ respectively. We note that total number density is necessarily a positive real number, i.e., $\rho_{0}=\rho_{\bm{\hat{n}}}+\rho_{\bm{-\hat{n}}} > 0$, while the net spin density can be any real number lying between zero and $p$.
The non zero values of $\bm{p}=p\bm{\hat{n}}$ corresponds to an electronic fluid from a magnetic reservoir with magnetization direction given by $\bm{\hat{n}}$. In particular $p=1$ and  
$0 < p < 1$ corresponds to a fully polarized and partially polarized electronic fluid which
is the case for magnetic reservoirs. This in turn implies that $\rho_{\bm{\hat{n}}} \ge 0$, $\rho_{-\bm{\hat{n}}} \ge 0$, and the $i$th($i\,\in [x,y,z]$) Cartesian component of polarization vector can be any real number lying between $+p$ and $-p$, i.e, $-p \le p_{i} \le +p$. We would use this fact in solving eqs.() in section.

The treatment of non-magnetic reservoirs is rather subtle in presence of spin-dependent scattering. Toward this end we note that in the incident density matrix if we put $p=0$ it describes the unpolarized incident fluid which is equivalent to averaging over quantization axis with $p=1$,i.e,
\begin{equation}
\frac{1}{4\pi}\int \varrho(E,\hat{\bm{n}})|_{\{p=1\}} d\hat{\bm{n}} =(1/2) (\rho_{0}{I}_{2}^{s}). 
\end{equation}
However for the scattered density matrix these two are not equivalent because scattered density matrix $\rho_{sc}$ is spin space entangled and putting $p=0$ amounts to neglecting the entanglement. Hence for the  scattered density matrix
the correct procedure to treat non-magnetic reservoirs is to start with a fully polarized incident fluid ($|\bm{p}|=p=1$, in eq.(\ref{eq:Res-1}) pointing along an arbitrary quantization direction specified by the polar($\theta$) and azimuthal
($\phi$) angle, calculate the scattered density matrix and then average over all quantization axis. 
This procedure takes into account the effect of {\bf spin-orbital entanglement} in the scattered density matrix is at the heart of giving rises to {\bf fractional charge} as we will see later.

\subsection{The incident density matrix in the leads}
\label{sub-sec-R-IDM}
Since the leads are scattering free adiabatic extension of reservoirs, therefore, leads just acts as waveguides which carry the electronic fluid from the reservoir towards the sample and vice versa. Here we first write down the density matrix for the incident electronic fluid from the reservoir into the ideal leads(the scattering free region in Fig.(2a)). 
We consider a generic situation where electronic fluid is
simultaneously incident from both left and right reservoir on the scattering region(circled region in Fig.~(2a)). Because of the absence of scattering in the leads the incident density matrix in the leads is,
\begin{equation}
 \rho_{in,\alpha\beta}=(1/2)[\rho_{in,\alpha,0}\otimes {I}_{2}^{s}+\bm{p}_{in,\alpha}\cdot\bm{\sigma}]\delta_{\alpha\beta}, \,\,\, \{\alpha,\beta\} \in \{L,R\}.
\label{eq:IDM-1}
\end{equation}
where $\rho_{in,\alpha,0} $ is ``scalar number density `` of electronic fluid in $\alpha$th lead, and $\bm{p}_{in,\alpha}=p_{in,\alpha}\hat{\bm{n}}_{\alpha} \equiv p_{in,\alpha} ( \cos(\phi_{\alpha})\sin(\theta_{\alpha}),\sin(\phi_{\alpha})\sin(\theta_{\alpha}),\cos(\theta_{\alpha}) )$ is ``spin density vector`` or polarization vector of the incident electronic fluid in $\alpha$th lead. The $(\theta_{\alpha},\phi_{\alpha})$ are polar and azimuthal angle which defines the direction of polarization vector $\bm{p}_{in,\alpha}=\sum_{i\in\{x,y,z\}}p_{in,\alpha,i}\hat{i}$ in laboratory frame(see Fig.(1)) in $\alpha$th lead or reservoir.
The above density matrix can be re-expressed as a direct product of matrices in the lead and electronic spaces as,
\begin{eqnarray}
\varrho_{in}=(1/2)\left[\left(\begin{array}{cc}
            \rho_{in,L,0} &  0 \\
            0 & \rho_{in,R,0}
            \end{array} \right)\otimes {I}_{2}^{s}+\sum_{i}\left(\begin{array}{cc}
             {p}_{in,L,i} &  0 \\
            0 & {p}_{in,R,i}
            \end{array} \right)\otimes\sigma_{i}\right] \label{eq:IDM-2} \\
=(1/2)[\rho_{in}\otimes {I}^{s}_{2}+\sum_{i \in\{x,y,z\}}{\rho}_{in,i}\otimes{\sigma}_{i}] \equiv (1/2)[\rho_{in} {I}_{2}^{s}+\bm{\rho}_{in}\cdot \bm{\sigma}].
\label{eq:IDM-3}
\end{eqnarray} 
Therefore the incident density matrix $\varrho_{in}$ (latex symbol ''varrho'') in left side of eq.(\ref{eq:IDM-2}) is composed of four scalar matrices $[\rho_{in},\rho_{in,x},\rho_{in,y},\rho_{in,z}]$(latex symbol ''rho'') on right side of eq.(\ref{eq:IDM-3}),which 
can be written in a compact form using scalar density matrix $\rho_{in,0}$(latex symbol:``rho``) and {\bf vectorial density matrix},
$\bm{\rho}$ as is done on the extreme right in eq.(\ref{eq:IDM-3}). 
The dimensionality of matrices $(\rho_{in}$, and ${\bm{\rho}}_{in}$ appearing in eqs.(\ref{eq:IDM-2}) and (\ref{eq:IDM-3})is $ 2\times 2$ because we are considering a two-terminal(lead) system, hence for $N$ terminal system these matrices would accordingly be $N \times N$ diagonal matrices. The important point is that our formulation can be generalized for systems with arbitrary number of terminals.
The elements of diagonal matrices $\rho_{in}$ and  $\bm{\rho}_{in}$ are real numbers and real vectors  as required by the hermiticity of full density density matrix. 

\begin{center}
\begin{figure}
\includegraphics[width=\linewidth,height=3.0in,angle=0,keepaspectratio]{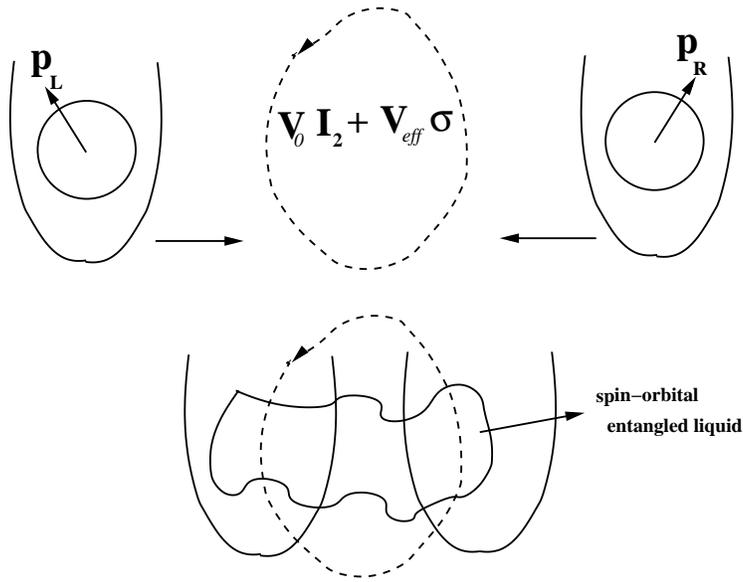}
\caption{\label{fig3.1} Upper Panel: The pictorial representation of incident density matrix given by eq.(\ref{eq:IDM-3}). Two component liquids from two reservoirs with uniform polarization vectors(the magnitude of polarization is independent of polar and azimuthal angles) , $\bm{p}_{L}$ and $\bm{p}_{R}$ approaches the spin-dependent potential. {Lower Panel:} Due to spin-dependent potential the initial density matrix evolves into final or scattered density matrix given by eq.(\ref{eq:SDM-3}) which is spin-space entangled liquid with a non-uniform polarization vector, i.e, magnitude of polarization is function of both lead indices as well the Cartesian indices. {\bf  From this picture we see that our formulation is equivalent to having two spin-independent potential wells separated by a region where potential is spin-dependent which is equivalent to having a spin-dependent potential step.}}
\end{figure}
\end{center}

The above incident density matrix encodes both {\bf equilibrium} as well {\bf non-equilibrium} conditions. 
First let us look at the {\bf equilibrium conditions}. Toward this end we note that physically the 
the leads are equivalent two potential wells which contains two liquids as shown in figure(2). The scattering region is equivalent to a spin-dependent potential step or barrier. Therefore connecting the two leads via scattering region is physically equivalent to having a two potential wells separated by a spin-dependent potential barrier. In other words
this systems is equivalent to a double well potential where potential barrier is spin-dependent and each potential well contains an initial density which is spin-space untangled. Now the question is, If this systems is left to itself and allowed to reach equilibrium  what is the final density matrix? The present formulation allows to calculate the final(scattered) state density matrix exactly and it turns out that the final or scattered  density matrix is {\bf spin-space entangled liquid} as we show in section(\ref{sec:SDM}).

The {\bf non-equilibrium} condition corresponds to a situation where incident density matrix is non-zero only in one of the leads, for example, suppose liquid is incident from left lead only then, $\rho_{in, R,l}=0, \, \forall \, l \in (0,x,y,z)$ in eq.(\ref{eq:IDM-2}) substituting this into final or scattered density matrix one would obtain the corresponding final state density matrix for non-equilibrium condition. The important point to note is that the final state density matrix will still be spin-space entangled and care should be taken while projecting the final state density matrix on the lead space. Therefore within the present formulation both equilibrium as well non-equilibrium condition are treated at equal footing. 

Now let us look at the physical observable associated with the incident density matrix(eqs.(\ref{eq:IDM-2}) and (\ref{eq:IDM-3}).
There are two physical observable, namely,  
number density which is a {\bf scalar observable} and spin density which is a {\bf vector observable}, associated with the incident electronic fluid described by the density matrix eqs.(\ref{eq:IDM-2}) and (\ref{eq:IDM-3}).  The total number density in incident fluid is a scalar observable and is given by the trace of density matrix(for details see the section \ref{S1-1} of supplementary information),
\begin{equation}
N_{in}=\text{Tr}(\varrho_{in})=\text{tr}(\rho_{in})=\rho_{in,L,0}+\rho_{in,R,0} \label{eq:IDM-4}
\end{equation}
while the total spin-density in the incident fluid is a vector observable and is given by the following trace,
\begin{eqnarray}
\bm{P}_{in}=\text{Tr}[\varrho_{in} \bm{\sigma}]&=&\text{tr}[{\bm{\rho}}_{in}] \nonumber \\
&=& \sum_{i}p_{in,L,i}\hat{i}+\sum_{i}p_{in,R,i}\hat{i} 
\equiv(\bm{p}_{in,L}+\bm{p}_{in,R}).\label{eq:IDM-5}
\end{eqnarray}
As we see from the above relation that total spin-density vector is a sum of two three dimensional vectors corresponding to the left and right reservoir. The generic initial condition where the left($\bm{p}_{in,L}$) and right ($\bm{p}_{in,R}$) are pointing along arbitrary direction in three dimensional space implies that the total spin-density vector also points along an arbitrary direction in three dimensional space. Geometrically these generic initial condition corresponds to a vector on sphere of radius $|(\bm{p}_{in,L}+\bm{p}_{in,R})|$.
The incident total polarization which is addition 
two three dimensional vectors lies on a sphere of radius $|(\bm{p}_{in,L}+\bm{p}_{in,R})|$, where polar and azimuthal angle of this vector defines the direction 
In particular if a reservoir is non-magnetic then the corresponding diagonal element of  $\bm{\varrho}$ is zero while the non-equilibrium initial condition in which electronic fluid is incident from only one of the reservoir then the corresponding diagonal element of scalar matrix $\rho_{in}$ as well the vectorial matrix  $\bm{\varrho}$ is zero. 
Therefore as discussed previously, the incident density matrix of eq.(\ref{eq:IDM-3}) is most generic one
as it encodes information about both magnetic as well non-magnetic reservoir at equal footing.

\section{Scattering Region: Scattering Potential and Scattering Matrix}
\label{sec-SPSM}
In this section we argue that most generic scattering potential for two component liquid has the form as given in the eq.(1) of introduction.
Toward this end we note that the incident electronic fluid described by the density matrix in eq.(\ref{eq:IDM-3}) meets various scattering centers as it approaches the elastic scattering region shown as circled area in Fig.(2a). 
These scattering centers are basically potential variations or discontinuities experienced by electronic fluid 
which can be sharp as well regular. The sharp potential discontinuities occurs at the interface or contact between ''leads and sample'' and at the ''sample boundaries'', while comparatively smooth potential variations occurs inside the sample due to presence of impurities: non-magnetic as well magnetic. It is important to remember that a magnetic impurity also acts as potential scatterer.
Therefore the total effective spin-independent potential,$V_{0,eff}$, due to various scattering centers discussed above is,
\begin{equation}
 V_{0,eff}=(V_{0,imp,NM}+V_{0,imp,M}+V_{0,contact}+V_{0,boundaries})\otimes \bm{I}_{2}^{s}
\label{eq:SP-pot0}
\end{equation}
where $V_{0,imp,NM}$ is scalar potential due to non-magnetic impurities, $V_{0,imp,M}$ is due to magnetic impurities, $V_{0,contact}$ is due to contacts and $V_{0,boundaries}$ arise due to sample boundaries.

The effective spin-dependent potential can arise due to spin-orbital coupling as well magnetic coupling. The spin-orbit coupling due to $V_{0,eff}$ is given by,
\begin{equation}
V_{SO}=\lambda_{so}(\bm{\nabla}V_{0,eff}(\bm{r}) \times  \bm{p}_{fluid})\cdot \bm{\sigma} 
\label{eq:SP-pot-so}
\end{equation}
where  $\bm{p}_{fluid}$ momentum vector of fluid,$\lambda_{so}$ is spin-orbit coupling strength and other symbols has usual meaning. The important point to realize that $V_{0,eff}$ has contribution from all scattering centers including magnetic scattering centers. Similarly the magnetic potential is given by,
\begin{equation}
V_{M}=\lambda_{M} (\sum_{i}\bm{m}_{i}) \cdot \bm{\sigma} \equiv \lambda_{M} \bm{M}_{I} \cdot \bm{\sigma}
\label{eq:SP-pot-mag}
\end{equation}
where $\lambda_{M}$ is magnetic coupling strength and $M_{I}$ is total magnetic moment vector of magnetic impurities.
Therefore the total scattering potential(spin-independent and spin-dependent) for the two component electronic fluid in the laboratory frame can be expressed as a sum of eqs.(\ref{eq:SP-pot0}),(\ref{eq:SP-pot-so}) and (\ref{eq:SP-pot-mag}) and leads to,
\begin{eqnarray}
V&=&V_{0,eff} {I}_{2}^{s}+ V_{SO}+V_{M}\equiv  V_{0,eff}\otimes  {I}_{2}^{s}+ \sum_{i\in\{x,y,z\}}V_{eff,i}\otimes \sigma_{i}\nonumber \\
&=&  V_{0} {I}_{2}^{s}+\bm{V}\cdot \bm{\sigma}
\label{eq:SP-1}
\end{eqnarray}
which is same as the eq.(1) of introduction.  In summary we have argued that effective scattering potential for two component electronic fluid
can be decomposed in spin space terms of a scalar potential $V_{0}$, and a vector potential $\bm{V}$ as is done eq.(\ref{eq:SP-1}) above. For sake of completeness we mention that inclusion of lead space turns $V_{0}$ into a $2\times 2$ scalar matrix and $\bm{V}$ into a $2\times 2$ vectorial matrix
similar to the incident density matrix given in eq.(\ref{eq:IDM-2}-\ref{eq:IDM-3}). Before we proceed further we would like to mention that the effective scattering potential defined in eq.(\ref{eq:SP-1}) is generic in the sense that we have not put symmetry constraints on it such as time reversal. In the rest of the paper we develop density matrix scattering theory for the generic scattering potential and consider time reversal only at the end of the paper in section(\ref{sec:time-rev}).

\subsection{Scattering Matrix:} 
\label{sec-SPSM1}
The generic scattering potential given in 
eq.(\ref{eq:SP-1}) causes both spin-independent as well spin-dependent scattering therefore an element $S_{\alpha\beta}$ ($\{\alpha,\beta\} \in \{L,R\}$) of scattering matrix $S$ can be decomposed in electronic space as, 
\begin{equation}
 S_{\alpha\beta}= S_{\alpha\beta,0} \otimes {I}^{s}_{2} + S_{\alpha\beta,x}\otimes \sigma_{x} +  S_{\alpha\beta,y}\otimes \sigma_{y} + S_{\alpha\beta,z}\otimes \sigma_{z}
\equiv S_{\alpha\beta,0} {I}^{s}_{2}+ {\bm{S}}_{\alpha\beta}\cdot \bm{\sigma},
\label{eq:SM-1}
\end{equation}
where complex scalar $(S_{\alpha\beta,0})$ and complex vector $\{\bm{S}_{\alpha\beta}=\sum_{i}S_{\alpha\beta,i}\hat{i}, i \in (x,y,z)\}$ are spin-independent and spin-dependent scattering amplitudes respectively. Moreover the complex scattering amplitude  $S_{\alpha\beta}$ defined above is a matrix hinting that it would give rise to {\bf non-abelian} phase shifts during the scattering processes. In-fact since spin-dependent scattering breaks the SU(2) symmetry in spin space therefore these scattering amplitudes will be a function of spin-quantization axes $\hat{\bm{n}}_{\alpha}$ and $\hat{\bm{n}}_{\beta}$ which defines the polarization directions in the leads, i.e,
\begin{eqnarray}
S_{\alpha\beta,l}=| S_{\alpha\beta,l}(\hat{\bm{n}}_{\alpha},\hat{\bm{n}}_{\beta})|exp[{\bm\dot{{\iota}}}\xi_{\alpha\beta,l}(\hat{\bm{n}}_{\alpha},\hat{\bm{n}}_{\beta})] \label{eq:polar-form}
\end{eqnarray}
where $| S_{\alpha\beta,l}(\hat{\bm{n}}_{\alpha},\hat{\bm{n}}_{\beta})|$ and $\xi_{\alpha\beta}(\hat{\bm{n}}_{\alpha},\hat{\bm{n}}_{\beta})$ are magnitude and phase of the complex scattering amplitude. In-fact even if leads are non-magnetic still the broken SU(2) symmetry in spin space affects the scattered density matrix in a non-trivial way therefore these angular correlation have to be taken into account properly. In the rest of the article we will omit the $(\hat{\bm{n}}_{\alpha},\hat{\bm{n}}_{\beta})$ for sake of simplicity and simply write the scattering amplitudes as
$S_{\alpha\beta,l}$.

The full scattering matrix ${\mathcal{S}}$(latex symbol ''mathcal``) in
the lead basis can be expressed as,
\begin{eqnarray}
\mathcal{S} &=&\left(\begin{array}{cc}
                      {S}_{LL} & {S}_{LR}\\
                      {S}_{RL} & {S}_{RR}
                      \end{array} \right) \nonumber \\
&=&\left(\begin{array}{cc}
                      {S_{LL,0}} & {{S}_{LR,0}} \\
                      {S_{RL,0}} & {S_{RR,0}}
                      \end{array} \right) \otimes {I}^{s}_{2}+\sum_{i\in\{x,y,z\}}\left(\begin{array}{cc}
                      {S}_{LL,i} & {S}_{LR,i} \\
                      {S}_{RL,i} & {S}_{RR,i}
                      \end{array} \right)\otimes{\sigma_{i}} \label{eq:SM-2}\\
&=&  {S}_{0}\otimes {I}^{s}_{2}+\sum_{i}S_{i}\otimes{\sigma_{i}} 
\equiv {S}_{0} {I}^{s}_{2}+\bm{S}\cdot\bm{\sigma}  \label{eq:SM-3}
\end{eqnarray}
where in Eqs.(\ref{eq:SM-2}-\ref{eq:SM-3}) above we have introduce $ 2 \times 2 $ dimensional matrices of complex numbers in the  space of particle $ \{{S}_{i},i\in(0,x,y,z)\}$. 
Further by introducing two dimensional spatial {\bf vectorial matrix} $\{\bm{S}=\sum_{i}{S}_{i}\hat{i},i\in(x,y,z)\}$,
we obtain the compact form, ${S}_{0} {I}^{s}_{2}+\bm{S}\cdot\bm{\sigma}$, on the rightmost side of eq.(\ref{eq:SM-3}),  which is similar to eq.(\ref{eq:IDM-3})for the incident density matrix. We note that the above scattering matrix has been obtained for the generic case without any symmetry constraints(constraints arising due to various symmetries will be discussed later). Using this scattering matrix in next section we calculate the scattered density matrix and obtain the expressions for {\bf charge density}(equivalently ``number density`` of scattered fluid) and {\bf spin density}(equivalently ``Spin Polarization'' of the scattered fluid) using which we discusses the conservation principle. 

\section{Scattered Density Matrix(SDM)}
\label{sec:SDM}
The incident two component electronic fluid described by the density matrix in eqs.(\ref{eq:IDM-2}) (equivalently (\ref{eq:IDM-3}) ) is scattered in the intermediate scattering region which gives rise to scattered fluid. The density matrix for the scattered fluid is completely known by the following relation,
\begin{eqnarray}
 \varrho_{sc}&=&{\mathcal{S}}\varrho_{in}{\mathcal{S}}^{\dagger} \nonumber \\
&=&[{S}_{0}{I}^{s}_{2}+\bm{S}\cdot\bm{\sigma}][\rho_{in} {I}^{s}_{2}+{\bm{\rho}}_{in}\cdot \bm{\sigma}][{S}_{0}^{\dagger}{I}^{s}_{2}+\bm{S}^{\dagger}\cdot\bm{\sigma}],\label{eq:SDM-1}
\end{eqnarray}
where we have made use of defining eqs.(\ref{eq:IDM-3}) and (\ref{eq:SM-3}) for incident density matrix($\varrho_{in}$) and scattering matrix ${\mathcal{S}}$ respectively. For sake of clarity we note that, $(\bm{S}\cdot\bm{\sigma})^{\dagger}=\sum_{i}(S_{i}\otimes\sigma_{i})^{\dagger}= 
\sum_{i}(S_{i}^{\dagger}\otimes\sigma_{i}^{\dagger})\equiv\bm{S}^{\dagger}\cdot\bm{\sigma} $,i.e., dagger operation for
direct product of matrices is distributive. To simplify the general expression (\ref{eq:SDM-1}), we make use of the identities (\ref{eq:S5}-\ref{eq:S9}) derived in appendix and obtain the
following simplified expression,(the details of simplification are given in supplementary information 1),
\begin{eqnarray}
\varrho_{sc}\hspace{-0.3cm}&=&\hspace{-0.3cm} [S_{0}\rho_{in}S_{0}^{\dagger}+\bm{S}\rho_{in}\cdot\bm{S}^{\dagger}+\{(\bm{S}\cdot{\bm{\rho}}_{in})S_{0}^{\dagger}+
S_{0}({\bm{\rho}}_{in}\cdot\bm{S}^{\dagger})\}
+\bm{\dot{\iota}}(\bm{S}\times{\bm{\rho}}_{in})\cdot \bm{S}^{\dagger}]\otimes {I}_{2} \nonumber \\
\hspace{-0.3cm}&+&\hspace{-0.3cm}[S_{0}{\bm{\rho}}_{in}S_{0}^{\dagger}+\{S_{0}\rho_{in}\bm{S}^{\dagger}
+\bm{S}\rho_{in}S_{0}^{\dagger}\}+\{(\bm{S}\cdot{\bm{\rho}}_{in})\bm{S}^{\dagger}
-(\bm{S}\times {\bm{\rho}}_{in})\times \bm{S}^{\dagger})\}]\cdot \bm{\sigma} \nonumber \\
\hspace{-0.3cm}&+&\hspace{-0.3cm}[\{\bm{\dot{\iota}}S_{0}({\bm{\rho}}_{in}\times \bm{S}^{\dagger})+\bm{\dot{\iota}}(\bm{S}\times{\bm{\rho}}_{in})S_{0}^{\dagger}\} 
+\bm{\dot{\iota}}(\bm{S}\rho_{in}\times \bm{S}^{\dagger}) ]\cdot\bm{\sigma}. \label{eq:SDM-2} \\
\hspace{-0.3cm}&=&\hspace{-0.3cm} \rho_{sc}\otimes {I}^{s}_{2}+\sum_{i}{\rho}_{sc,i}\otimes{\sigma}_{i} \hat{i}\equiv \rho_{sc}\otimes {I}^{s}_{2}+\bm{\rho}_{sc}\cdot \bm{\sigma}\label{eq:SDM-3}.
\end{eqnarray}
The hermiticity of scattered density matrix is evident in the compact expression (\ref{eq:SDM-1}) while in the simplified expression (\ref{eq:SDM-2}) we have combined terms which are together hermitian inside a curly bracket and terms which are individually hermitian have been 
shown without any bracket around them.
The detailed proof of hermiticity of various terms is given in the Supplementary Information(\ref{sec-S3}). 

Now we stress an important point. First we notice that the scattered density matrix $\varrho_{sc}$(eq.(\ref{eq:SDM-3})) is composed of a scalar density matrix $\rho_{sc}$, and vectorial density matrix $\bm{\rho}_{sc}$ respectively, which is similar to the structure of incident density matrix eq.(\ref{eq:IDM-3}) and scattering matrix eq.(\ref{eq:SDM-3}). The important difference between incident(\ref{eq:IDM-3}) and scattered density matrix(\ref{eq:SDM-3}) is that while incident density matrix diagonal in spatial(lead+Cartesian indices) and spin space,i.e., it is separable in spatial and spin space. In other words incident density matrix is un-entangled in spatial and spin space. However the scattered density matrix is non-separable in spatial and spin space because it is neither diagonal in spin space nor in lead space. This is because the scattered density matrix($\varrho_{sc}$) in eq.((\ref{eq:SDM-3})) is {\bf spin-space(orbital) entangled} because both $\rho_{sc}$ and $\bm{\rho}_{sc}$ in eq.(\ref{eq:SDM-3}) depends on the spatial coordinates as they are function of lead index($\alpha, \beta$) as well of the Cartesian indices $i,j,k$, hence the scattered density matrix($\varrho_{sc}$) in eq.((\ref{eq:SDM-3})) can not be written as product form in spatial and spin space.
The non-diagonality of scalar part of scattered density matrix $\rho_{sc}$ in spin space which arise due to {\bf spin-orbital} entanglement is at the heart of giving rise to {\bf fractional charge} and intimately related to the {\bf non-abelian phase} of scattering matrix as we will see later.

\section{Charge and Spin density in scattered fluid}
\label{sec-Num-DP}
The charge density or equivalently ``number density``  of scattered fluid is given by $\text{Tr}(\rho_{sc})$ which has been calculated explicitly in supplementary information (\ref{sec-S4} eq.(\ref{eq:S28-1})

The expression for charge density for the case when the reservoirs are magnetic is given by,
\begin{eqnarray}
\text{Tr}( \varrho_{sc})= 2\text{tr} (\rho_{sc})=2\sum_{\alpha}(\rho_{sc})_{\alpha\alpha}
= 2 \sum_{\alpha\beta}\left[{\cal{C}}^{\alpha\beta} \rho_{in,\beta,0}  + \sum_{(ijk)_{c}}2{\cal{Q}}^{\alpha\beta}_{0ijk}p_{in,\beta,i}\right],\label{eq:DP-1} \\
=2 \sum_{\alpha\beta}\left[{\cal{C}}^{\alpha\beta} \rho_{in,\beta,0} +2{\cal{Q}}^{\alpha\beta}_{0xyz}p_{in,\beta,x}+2{\cal{Q}}^{\alpha\beta}_{0yzx}p_{in,\beta,y}+2{\cal{Q}}^{\alpha\beta}_{0zxy}p_{in,\beta,z}\right] \label{eq:DP-1.1}
\end{eqnarray}
where $\{\alpha , \beta \} \in \{ L, R \} $ corresponding to left and right lead and the summation symbol $(i j k)_{c}$ implies that this summation is only over cyclic permutation of Cartesian axis, 
i.e.,  $(i j k)_{c}\in \{(x,y,z),(y,z,x), (z, x, y)\}$. To avoid confusion we have written this summation explicitly in eq. (\ref{eq:DP-1.1}).
The coefficients ${\cal{C}}$'s and ${\cal{Q}}$'s appearing in the expressions (\ref{eq:DP-1}) are defined as, 
\begin{eqnarray}
{\cal{C}}^{\alpha\beta}&=&(|S_{\alpha \beta,0}|^{2} +\sum_{i}|S_{\alpha \beta,i}|^{2}) \label{eq:DP-C1} \\
{\cal{Q}}_{0ijk}^{\alpha\beta}&=&[\text{Re}(S_{\alpha \beta,0}S_{\alpha \beta,i}^{*})-
\text{Im}(S_{\alpha\beta,j}^{*} S_{\alpha\beta,k})],\label{eq:DP-Q1}
\end{eqnarray}
The coefficients, ${\cal{C}}^{\alpha\beta} > 0$, as it is sum of spin-independent and spin-dependent scattering probabilities in this sense these are {\it classical coefficient}. On the other hand the coefficients, ${\cal{Q}}_{0ijk}^{\alpha\beta}$ are {\it quantum coefficients}
because they arises due to quantum interference between various components of scattering matrix element, ${\cal{S}}_{\alpha\beta}=S_{\alpha\beta,0} \bm{I}_{2}^{s}+ \bm{S}_{\alpha\beta} \cdot \bm{\sigma}$, defined in eq.(\ref{eq:SM-1}). The {\it quantum coefficients} can be put into a vector form
\begin{eqnarray}
\bm{Q}^{\alpha\beta}&=&\sum_{(ijk)_{c}}{\cal{Q}}^{\alpha\beta}_{0ijk}\hat{i} \nonumber \\
&\equiv& [{\cal{Q}}^{\alpha\beta}_{0xyz}\hat{x}+{\cal{Q}}^{\alpha\beta}_{0yzx}\hat{y}+{\cal{Q}}^{\alpha\beta}_{0zxy}\hat{z}] \label{eq:DP-VQ1}
\end{eqnarray}
and writing the polarization vector of incident fluid in lead $\beta$ as, 
\begin{eqnarray}
\bm{p}_{in,\beta}=p_{in,\beta}{\hat{\bm{n}}}_{\beta}&\equiv& p_{in,\beta}[\cos{\phi}_{\beta}\sin{\theta}_{\beta}\hat{x}+\sin{\phi}_{\beta}\sin{\theta}_{\beta}\hat{y}+\cos{\theta}_{\beta}\hat{z}] \\
&\equiv& p_{in,\beta}\sum_{i} n_{\beta,i}\hat{i} \label{eq:pol-in-vec}
\end{eqnarray}
which allows us to expresses the eq.(\ref{eq:DP-1}) which gives the total number density, $N_{sc}=\text{Tr}( \varrho_{sc})$ in the scattered fluid as,
\begin{eqnarray}
N_{sc}&=&2 \sum_{\alpha\beta}\left[ {\cal{C}}^{\alpha\beta} \rho_{in,\beta,0}  + 2 \bm{Q}^{\alpha\beta} \cdot {\hat{\bm{n}}}_{\beta} p_{in,\beta} \right ]\label{eq:DP-1.2} \\
&\equiv& \sum_{\alpha}[N_{sc,\alpha, \beta=L}+N_{sc,\alpha, \beta=R}] \label{eq:DP-1.3}
\end{eqnarray}
The expression(\ref{eq:DP-1.2}) gives the total number density in the scattered fluid which can be written as a sum of scattered density in left lead and right lead as is done in the expression (\ref{eq:DP-1.3}). However the important point to note is that number density in left or right lead also depends on the vector $\bm{Q}^{\alpha\beta}$ which arises due to quantum interference and plays an important role as it encodes the information about {\bf non-abelian} scattering phase which will be discussed in the next section.

Before we proceed further we note that these same quantum interference coefficients also appear in the spin density vector,$\bm{P}_{sc}=\text{Tr}(\rho_{sc}\bm{\sigma})=2\text{tr}[{\bm{\varrho}}_{c}]$, of scattered fluid, which has been calculated in  supplementary information (\ref{sec-S4} eq.(\ref{eq:S28-1}) and is given by,
\begin{eqnarray}
\bm{P}_{sc}&=& \sum_{(ijk)_{c},\alpha\beta}2\left\{[\text{Re}(S_{\alpha\beta,0}S_{i,\alpha\beta}^{*})-\text{Im}(S_{\alpha\beta,j}S_{\alpha\beta,k}^{*})]\hat{i}\right\}\rho_{in,\beta,0}\nonumber \\
&+&\sum_{(ijk)_{c},\alpha\beta}[|S_{\alpha\beta,0}|^{2}+ |S_{\alpha\beta,i}|^{2}- |S_{\alpha\beta,j}|^{2}-|S_{\alpha\beta,k}|^{2}] p_{in,\beta,i}\hat{i} \nonumber \\
&+&\sum_{(ijk)_{c},\alpha\beta}2[\text{Re}(S_{\alpha\beta,i}S_{\alpha\beta,j}^{*})-\text{Im}(S_{\alpha\beta,0}S_{\alpha\beta,k}^{*})]p_{in,\beta,j}\hat{i} \nonumber \\
&+&\sum_{(ijk)_{c},\alpha\beta}2[\text{Re}(S_{\alpha\beta,i}S_{\alpha\beta,k}^{*})-\text{Im}(S_{\alpha\beta,0}^{*}S_{\alpha\beta,j})]p_{in,\beta,k}\hat{i} \label{eq:DP-SD}
\end{eqnarray}
We notice the coefficient of first term is nothing but the vector $\bm{Q}^{\alpha\beta}=\sum_{(ijk)_{c}}{\cal{Q}}^{\alpha\beta}_{0ijk}$ which appears in the charge density expressions (\ref{eq:DP-1}) and (\ref{eq:DP-1.2}). 
The similar but different quantum interference such as ${\cal{Q}}^{\alpha\beta}_{ijk0}$ and ${\cal{Q}}^{\alpha\beta}_{ik0j}$ which multiplies  $n_{\beta,j}$ and $n_{\beta,k}$ also appear in the above expression(\ref{eq:DP-SD}). Therefore in the charge density (\ref{eq:DP-1}) and spin-density vector (\ref{eq:DP-SD}) all possible combination of {\it quantum interference} terms appears which has to be so because all these terms are at equal footing as far as {\it quantum interference} effects are concerned.

The expression (\ref{eq:DP-SD}) can also be written in terms of dot product of two vector by defining a vector $\bm{P}^{\alpha\beta}$ as,
\begin{eqnarray}
\bm{P}^{\alpha\beta}&=&\sum_{(ijk)_{c}}\Bigg\{[|S_{\alpha\beta,0}|^{2}+ |S_{\alpha\beta,i}|^{2}- |S_{\alpha\beta,j}|^{2}-|S_{\alpha\beta,k}|^{2}]n_{\beta,i}\nonumber \\ &+&2[\text{Re}(S_{\alpha\beta,i}S_{\alpha\beta,j}^{*})-\text{Im}(S_{\alpha\beta,0}S_{\alpha\beta,k}^{*})]n_{\beta,j}\nonumber \\
&+&2[\text{Re}(S_{\alpha\beta,i}S_{\alpha\beta,k}^{*})-\text{Im}(S_{\alpha\beta,0}^{*}S_{\alpha\beta,j})]n_{\beta,k}\Bigg\}\hat{i}\label{eq:DP-SD1}
\end{eqnarray}
where we have made use of the expression ({\ref{eq:pol-in-vec}}) for the incident polarization vector.
Therefore the expression (\ref{eq:DP-SD}) can be expressed in a compact form as,
\begin{equation}
\bm{P}_{sc}=\sum_{\alpha\beta} [\bm{Q}^{\alpha\beta}\rho_{in,\beta,0}+ \bm{P}^{\alpha\beta} p_{in,\beta}] \label{eq:DP-SD-1} 
\end{equation}
The expression (\ref{eq:DP-1.2}) and (\ref{eq:DP-SD-1}) defines the two observables for the scattered fluid, namely, the scalar observable-- Number Density $N_{sc}$ and the vectorial observable spin density vector $\bm{P}_{sc}$. In presence of spin-dependent scattering it is expected that spin density vector is not a conserved vector which is explicitly seen from the expression for $\bm{P}_{sc}$ eq.(\ref{eq:DP-SD-1}) which is not same as that of incident spin-density vector $\bm{P}_{sc}$, given by expression (\ref{eq:IDM-5}). We will discusses the conservation laws in some details in section (6). 

For the time being let us try to visualize the vectors $\bm{P}_{in}$ and $\bm{P}_{sc}$ in terms of polar coordinates ($\theta, \phi$) which would give us a physical picture. Towards this end we note that incident spin-density is a vector of constant magnitude, therefore, it defines a uniform Bloch sphere of radius $|\bm{P}_{in}|=|\bm{p}_{in,L}+\bm{p}_{in,R}|$ which is shown in the left panel of the Fig.(4). This Uniform Bloch sphere corresponds to having a constant(uniform) number and spin density,i.e, $N_{in} =\rho_{in,L,0}+\rho_{in,R,0}$ and $|\bm{P}_{in}|$ . In other words every point in the uniform Bloch sphere gets defined by only two variables ,i.e, polar and azimuthal angle, because both $N_{in}$ and $|\bm{P}_{in}|$ which are fixed constant numbers throughout the sphere and do not depend on the 
polar and azimuthal angles. However the magnitude of scattered spin density vector is not constant rather it is a function of polar and azimuthal angles,i.e, $|\bm{P}_{sc}|=F(\theta,\phi)$, because the vector $\bm{P}_{sc}$ depends on vectors $\bm{Q}^{\alpha\beta}$, $\bm{P}^{\alpha\beta}$ which are not vectors of constant magnitude.   
Similarly the scattered number density $N_{sc}$ given by expression (\ref{eq:DP-1.2}) is also a function of $\bm{Q}^{\alpha\beta}$,therefore, it would also be a function of polar coordinates,i.e,$N_{sc}=G(\theta,\phi)$. Therefore the geometrical object corresponding to vector $\bm{P}_{sc}$
would be a a generalized three dimensional object with variable radius given by a function $F(\theta,\phi)=|\bm{P}_{sc}|$. Hence each point in and on this spherical object is associated with five variables, namely, $N_{sc}=G(\theta,\phi)$ and $|\bm{P}_{sc}|=F(\theta,\phi)$ and the unit vector
$\hat{\bm{P}}_{sc}$ which defines the local spin-quantization axis on this object. This is equivalent to associating a scalar function $N_{sc}$ and a vector function $\bm{P}_{sc}$ to each point in and on this three dimensional object. Physically this implies that each point in and on this three dimensional object gets defined by the five functions of polar and azimuthal angles, namely, a ''local number density($(N_{sc}(\theta,\phi)$)`` and ``local spin-density($|\bm{P_{sc}}|(\theta,\phi)$'' which is pointing along an direction defined by unit vector $\hat{\bm{P}}_{sc}(\theta,\phi)$. Moreover $N_{sc}$ and $|\bm{P_{sc}}|$ are note independent functions implying that {\bf spin degree} and {\bf spatial degrees} of freedom are entangled. This {\bf spin-space entanglement} can be appreciated in a physical was if we look at the illustrative Bloch spheres in Fig.(3) ( a cartoon for the purpose of illustration only) for the incident(left panel) and scattered density(middle panel) where for the incident fluid the Bloch sphere is a uniform sphere with constant radius while for the scattered density it is en ''Entangled Bloch Sphere''. Here for the purpose of illustration we have plotted
a simple function $1+\sin[5\phi]\sin[10\theta]/10$, in general this ''Entangled Bloch Sphere'' will be a more complicated three dimensional topological object. The right most panel show the superposition of the left and the middle figure where we can see that initial uniform sphere has evolved into non-uniform sphere. 

\begin{center}
\begin{figure}
\includegraphics[width=\linewidth,height=3.0in,angle=0,keepaspectratio]{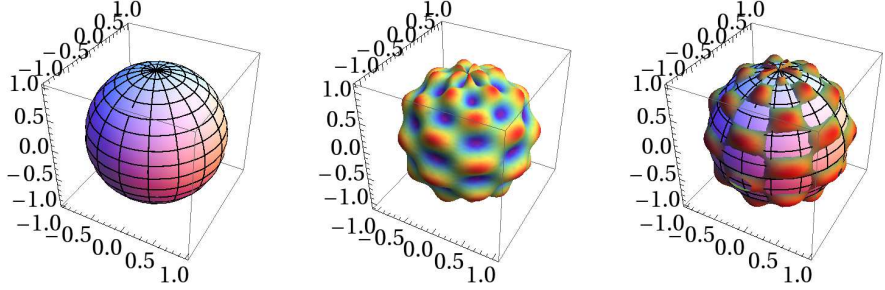}
\caption{\label{fig4} Illustrative figure(courtesy Mathematica): {\bf Left Panel:} Uniform Bloch Sphere for the incident fluid. It has constant radius and no surface modulation representative of the fact that ''number density'' as well ''spin-density'' are uniform throughout the sphere. {\bf Middle Panel:} Showing ''Entangled Bloch Sphere'' where each point inside and on the surface of sphere is associated with two observables a scalar number density $N_{sc}$ and a vector spin-density $P_{sc}$ given by eq.(\ref{eq:DP-1.2}) and (\ref{eq:DP-SD-1}), respectively which varies from in this three dimensional spherical surface. In general this ''Entangle Bloch Sphere'' will be a more complicated three dimensional topological object here for the purpose of illustration we have simply plotted the function $1+\sin[5\phi]\sin[10\theta]/10$.} {\bf Right Panel:}
Superposition of left and right figures showing the difference between the initial and final Bloch Spheres.
\end{figure}
\end{center}

From this discussion we see that the ''local number density`` need not to be an integer(in units of electronic charge $e$) and ``local spin-density'' also need not be half integer(Fermions) or integer(Bosons). In fact what we have described is a ``Topological quasi particle'' which is nothing but an effective collective excitation with 
fractional quantum numbers,i.e, fractional charge\cite{Klitzing,Tsui,laughlin,Jain1,Jain2,Jain3,Nayak,Hasan,Wen} and fractional spin also know as ``topological spin''\cite{Wen1,gur,Nread}. This fractionalization of quantum numbers would be further discussed in a mathematically rigorous way in section () from the point of view of conservation laws.

From the discussion above it clear that $N_{sc}$ and $P_{sc}$ are entangled reflecting the {\bf spin-space entanglement} or more commonly {\bf spin-orbital entanglement}. This is also seen in the charge density expression,eq.(\ref{eq:DP-1}), ${\cal{Q}}_{0ijk}^{\alpha\beta}$ multiplies with the polarization components,$p_{in,\beta,i}$ of incident fluids while for the spin density expression eq.(\ref{eq:DP-1}), ${\cal{Q}}_{0ijk}^{\alpha\beta}$ multiplies with the scalar number density $\rho_{in,\beta,0}$ of incident fluid. Therefore in the scattered fluid {\bf charge(number)} and {\bf spin} degree are entangled. If one assumes that ${\cal{Q}}_{0ijk}^{\alpha\beta}=0$ this would amount to neglecting the {\bf spin space entanglement}. Because of this reason the case of non-magnetic reservoirs(incident fluid being un-polarized) is not equivalent to putting $\bm{p}_{in,\beta} =0 $ in eq.(\ref{eq:DP-1.2}) for the scattered fluid. Rather for the scattered fluid one should take $|\bm{p}_{in,\beta}|=1$ and average over all possible quantization axis which corresponds to having the incident fluid un-polarized. If both the reservoirs are un-polarized then performing the integration over ${\bm{n}}_{\alpha}$ and  ${\bm{n}}_{\beta}$ in eq.(\ref{eq:DP-1}) leads to. 
\begin{eqnarray}
 \frac{1}{4\pi}\int \text{Tr}(\varrho_{sc})d\hat{\bm{n}}_{\alpha}d\hat{\bm{n}}_{\beta}&=&2[ \sum_{\alpha\beta}\frac{1}{4\pi}\int{\cal{C}}^{\alpha\beta}d\hat{\bm{n}}_{\alpha}d\hat{\bm{n}}_{\beta}] \rho_{in,\beta,0}\nonumber\\
&+&\frac{1}{4\pi}[\sum_{(ijk)_{c},\alpha\beta}\int 2{\cal{Q}}^{\alpha\beta}_{0ijk}d\hat{\bm{n}}_{\alpha}d\hat{\bm{n}}_{\beta}]. \label{eq:DP-NM}
\end{eqnarray}
From the above expression we see that if ${\cal{Q}}^{\alpha\beta}_{0ijk}$ is taken equal to zero it would amount to neglecting the spin-space entanglement. This aspect is closely related to the broken SU(2) invariance also as we discusses in the next subsection below.

\section{Broken SU(2) invariance for scattered fluid}
\label{sec-SU2-inv}
If the incident fluid is polarized, i.e, reservoirs are magnetic than SU(2) invariance is broken explicitly even for the incident fluid and this would naturally reflect in the scattered fluid. However for sake of completeness let us write the scattered charge density for a fixed lead $\beta$ which follows from expression (\ref{eq:DP-1.2}) and is given by,
\begin{equation}
N_{sc,\beta,0}(\rho_{in,\beta,0},{p}_{in,\beta},{\hat{\bm{n}}}_{\beta}) =[\sum_{\alpha}{\cal{C}}^{\alpha\beta}] \rho_{in,\beta,0}  \pm \sum_{\alpha}2  | \bm{Q}^{\alpha\beta}\cdot {\hat{\bm{n}}}_{\beta}| {p}_{in,\beta} 
\end{equation}
For polarized incident fluid ${p}_{in,\beta}\neq 0$, therefore, the scattered charge density for opposite direction $\pm{\hat{\bm{n}}}_{\beta}$ is not equal,i.e,
\begin{equation}
N_{sc,\beta,0}(\rho_{in,\beta,0},{p}_{in,\beta}, +{\hat{\bm{n}}}_{\beta}) \neq N_{sc,\beta,0}(\rho_{in,\beta,0},{p}_{in,\beta},-{\hat{\bm{n}}}_{\beta}) 
\end{equation}
which is explicitly shows that SU(2) invariance is broken for scattered charge density if the incident fluid is polarized. In fact for fixed $\rho_{in,\beta,0}$ and ${p}_{in,\beta}$ if $N_{sc,\beta,0}(+{\hat{\bm{n}}}_{\beta})$ is taken equal to $N_{sc,\beta,0}( -{\hat{\bm{n}}}_{\beta})$
which is possible only if coefficients ${\cal{Q}}^{\alpha\beta}_{0ijk}$ are zero. This would imply that the conductance of a two terminal magnetic system, which is proportional scattered Number density, would be independent of magnitude of magnetization (see the first equation) as well independent of which direction the magnetization is pointing. In other words it would imply complete absence of tunneling magneto resistance\cite{Moodera-TMR} as well anisotropic magneto-resistance\cite{Mcguire-AMR,Tribhuvan-PRB-70,Tribhuvan-PRB-75} which is unphysical result. Therefore the coefficients ${\cal{Q}}^{\alpha\beta}_{0ijk}$ are necessarily non-zero. 

The case of un-polarized incident fluid corresponding to non-magnetic reservoir can be dealt the way it was done to obtain the expression (\ref{eq:DP-NM}). Here we discusses another equivalent way to treat the non-magnetic reservoirs. 
Toward this end we note that an un-polarized liquid from a non-magnetic reservoir can be described as composed of incoherent mixtures of two fully polarized liquids along opposite direction, i.e, along $\pm{\hat{\bm{n}}}_{\beta}$ direction with ${p}_{in,\beta}=1$, which is equivalent to saying that incident density matrix is in completely mixed state. Mathematically this means that incident density matrix is SU(2) invariant. 
Now let us look at the charge density in the scattered fluid for these two components of the incident mixtures which is given by,
\begin{eqnarray}
N_{sc,\beta,0}(\rho_{in,\beta,0},1, +{\hat{\bm{n}}}_{\beta}) =[\sum_{\alpha}{\cal{C}}^{\alpha\beta}] \rho_{in,\beta,0}  \pm \sum_{\alpha}2  | \bm{Q}^{\alpha\beta}\cdot {\hat{\bm{n}}}_{\beta}|  \\
N_{sc,\beta,0}(\rho_{in,\beta,0},1, -{\hat{\bm{n}}}_{\beta}) =[\sum_{\alpha}{\cal{C}}^{\alpha\beta}] \rho_{in,\beta,0}  \mp \sum_{\alpha}2  | \bm{Q}^{\alpha\beta}\cdot {\hat{\bm{n}}}_{\beta}| 
\end{eqnarray}
Now if we impose SU(2) invariance for the scattered density matrix, than it requires that above two densities corresponding to $\pm{\hat{\bm{n}}}_{\beta}$ in the scattered fluid should be same, which is possible only if,
\begin{equation}
 |\bm{Q}^{\alpha\beta}\cdot {\hat{\bm{n}}}_{\beta}|=0
\end{equation}
The above condition should be satisfied for arbitrary $\theta_{\beta}$ and $\phi_{\beta}$ defining the quantization axis, ${\hat{\bm{n}}}_{\beta}=(\sin(\theta_{\beta})\cos(\phi_{\beta}), \sin(\theta_{\beta})\sin(\phi_{\beta}), \cos(\theta_{\beta}))$,
which in general is vector of three non zero real numbers. Therefore the above condition can only be satisfied if the vector $\bm{Q}^{\alpha\beta}=0$ which would amount to neglecting the spin-space entanglement as discussed above. 
Therefore we have shown the broken SU(2) invariance for the scattered density matrix requires  vector $\bm{Q}^{\alpha\beta}\neq 0$

In the next section using the coefficients  ${\cal{Q}}_{0ijk}^{\alpha\beta}$ defined in eq.(\ref{eq:DP-Q1}) and the corresponding vector $\bm{Q}^{\alpha\beta}$ defined in eq.(\ref{eq:DP-VQ1}) we show in a mathematically rigorous way that non-zero values of these requires that scattering matrix is {\bf non-unitary} which is equivalent to having {\bf non-abelian} scattering phases. 
In short it is time to start a mathematical journey from {\bf Abelian(Unitary)} hydrodynamics
to {\bf Non-Abelian(Non-Unitary)} hydrodynamics equivalently to {\bf spin-orbital entangled(Anyonic)} liquids.

\section{Abelian(Unitary) and Non-Abelian(Non-Unitary) Hydrodynamics}
\label{AB-NAB}
As discussed above the {\bf quantum interference coefficients}, $Q_{0ijk}^{\alpha\beta}$, plays an important role in determining whether the dynamics is Abelian and Unitary or Non-Abelian and Non-Unitary. Toward this end we consider complex scattering matrix element ${\cal{S}}_{\alpha\beta} =(S_{\alpha\beta,0} \bm{I}_{2}+ \bm{S}_{\alpha\beta} \cdot \bm{\sigma})$, which themselves are $2 \times 2$ matrices and note that coefficients ${\cal{Q}}^{\alpha\beta}_{0ijk}$ are related to these scattering matrix elements in the following way,
\begin{eqnarray}
{\cal{S}}_{\alpha\beta} {\cal{S}}^{*}_{\alpha\beta}&=&(S_{\alpha\beta,0} \bm{I}_{2}+ \bm{S}_{\alpha\beta} \cdot \bm{\sigma})(S_{\alpha\beta,0} ^{*}\bm{I}_{2}+ \bm{S}_{\alpha\beta}^{*} \cdot \bm{\sigma}) \nonumber \\
&=& (|S_{\alpha\beta,0}|^{2}+\sum_{i}|S_{i,\alpha \beta}|^{2})\bm{I}_{2}^{s}+[(S_{\alpha\beta,0}\bm{S}_{\alpha\beta}^{*}+S_{\alpha\beta,0}^{*}\bm{S}_{\alpha\beta})+\bm{\dot{\iota}}(\bm{S}_{\alpha\beta} \times \bm{S}_{\alpha\beta}^{*})]\cdot \bm{\sigma} \nonumber \\
&=&{\cal{C}}^{\alpha\beta} \bm{I}_{2}^{s}+[\sum_{i} 2\text{Re}(S_{\alpha\beta,0}{S}_{\alpha\beta,i}^{*})+\bm{\dot{\iota}}\sum_{(ijk)_{c}}(S_{\alpha\beta,j}S_{\alpha\beta ,k}^{*}-S_{\alpha\beta,k}S_{\alpha\beta,j}^{*})]\sigma_{i} \nonumber \\
&=&{\cal{C}}^{\alpha\beta} \bm{I}_{2}^{s}+\sum_{(ijk)_{c}}2[\text{Re}(S_{\alpha\beta,0}{S}_{\alpha\beta,i}^{*}-\text{Im}(S_{\alpha\beta,j}S_{\alpha\beta,k}^{*})]\sigma_{i} \label{eq:AB-NAB-1} \\
&=& {\cal{C}}^{\alpha\beta} \bm{I}_{2}^{s}+\sum_{(ijk)_{c}}2{\cal{Q}}^{\alpha\beta}_{0ijk}\sigma_{i} 
%\equiv \text{tr}[A_{0}]+\sum_{i}\text{tr}[A_{i}]\sigma_{i} \label{eq:AB-NAB-2}
\end{eqnarray}
Now to discusses under what condition coefficients ${\cal{Q}}^{\alpha\beta}_{0ijk}$ can be zero or non-zero 
we use polar form for the complex scattering amplitudes ${\cal{S}}_{\alpha\beta}$ defined in
eq.(\ref{eq:polar-form}) which leads to the following explicit expression, 
for the coefficients ${\cal{Q}}^{\alpha\beta}_{0ijk}$ appearing in eq.(\ref{eq:DP-1})(equivalently in eq.(\ref{eq:DP-1.1})), 
\begin{eqnarray}
{\cal{Q}}_{0xyz}^{\alpha\beta}=|S_{\alpha\beta,0}||S_{x,\alpha\beta}|\cos(\xi_{\alpha\beta,0}-\xi_{\alpha\beta,x})-|S_{\alpha\beta,y}||S_{\alpha\beta,z}|\sin(\xi_{\alpha\beta,y}-\xi_{\alpha\beta,z}) \label{eq:AB-NAB-x}\\
{\cal{Q}}_{0yzx}^{\alpha\beta}=|S_{\alpha\beta,0}||S_{\alpha\beta,y}|\cos(\xi_{\alpha\beta,0}-\xi_{\alpha\beta,y})-|S_{\alpha\beta,z}||S_{\alpha\beta,x}|\sin(\xi_{\alpha\beta,z}-\xi_{\alpha\beta,x}) \label{eq:AB-NAB-y}\\
{\cal{Q}}_{0zxy}^{\alpha\beta}=|S_{\alpha\beta,0}||S_{\alpha\beta,z}|\cos(\xi_{\alpha\beta,0}-\xi_{\alpha\beta,z})-|S_{\alpha\beta,x}||S_{\alpha\beta,y}|\sin(\xi_{\alpha\beta,x}-\xi_{\alpha\beta,y}).\label{eq:AB-NAB-z}
\end{eqnarray}
Here it is implicit that $|S_{\alpha\beta,l}|\equiv |S_{\alpha\beta,l}(\hat{\bm{n}}_{\alpha},\hat{\bm{n}}_{\beta})|$, $\xi_{\alpha\beta,l}\equiv \xi_{\alpha\beta,l}(\hat{\bm{n}}_{\alpha},\hat{\bm{n}}_{\beta})$, where $\hat{\bm{n}}_{\alpha}$ and $\hat{\bm{n}}_{\beta}$ are unit vectors defining the polarization direction in the leads see section(\ref{sub-sec-R-IDM}). 

It is easy to see that the above three expression are zero if spin-dependent scattering is absent which implies, $|S_{\alpha\beta,l}|=0, \forall \,\, l \in \{0,x,y,z\}$. However in presence of spin-dependent scattering,i.e, $|S_{l,\alpha\beta}|\neq 0 $, if one demands that ${\cal{Q}}^{\alpha\beta}_{0ijk}$ are zero, which can only be satisfied if one assume that phase of spin-dependent scattering amplitudes are all same,i.e, $\xi_{\alpha\beta,x}=\xi_{\alpha\beta,y}=\xi_{\alpha\beta,z}=\xi_{\alpha\beta}$ which in-turn implies that
\begin{eqnarray}
\cos(\xi_{\alpha\beta,0}-\xi_{\alpha\beta})=0 \implies (\xi_{\alpha\beta,0}-\xi_{\alpha\beta})=(\pi/2)+ n \pi
\label{uni-phase}
\end{eqnarray}
where $n$ is an arbitrarily chosen fixed integer. Using the above constraint on scattering phases let us rewrite scattering matrix element as follows,
\begin{eqnarray}
{\cal{S}}_{\alpha\beta}&=&S_{\alpha\beta,0} \bm{I}_{2}+ \bm{S}_{\alpha\beta} \cdot \bm{\sigma} \nonumber \\
&=& \exp(\bm{\dot{\iota}}\xi_{\alpha\beta,0})\left[|S_{\alpha\beta,0}|\bm{I}_{2}^{s}+\exp[\bm{\dot{\iota}}(\xi_{\alpha\beta,0}-\xi_{\alpha\beta})] \sum_{i}|S_{\alpha\beta,i}|\sigma_{i}\right]\nonumber \\
&=&\exp(\bm{\dot{\iota}}\xi_{\alpha\beta,0})\left[|S_{\alpha\beta,0}|\bm{I}_{2}^{s}+\bm{\dot{\iota}} \sum_{i}|S_{\alpha\beta,i}|\sigma_{i}\right]\nonumber \\
&\equiv&|S_{\alpha\beta,0}|\bm{I}_{2}^{s}+\bm{\dot{\iota}} \sum_{i}|S_{\alpha\beta,i}|\sigma_{i} \label{quaternion}
\end{eqnarray}
where in the last line above we have removed the overall phase $\exp(\bm{\dot{\iota}}\xi_{0,\alpha\beta})$. Therefore an element of scattering $ {\cal{S}}_{\alpha\beta}$ which originally had four independent phases as it was composed of four complex numbers, one four spin-independent scattering and three for spin-dependent scattering, has been reduced to a {\bf real quaternion} in eq.(\ref{quaternion})\cite{Dyson-Mehta,Altland,RM-Beena} which 
is composed of four real number and a fixed relative phase between spin-independent and spin-dependent scattering amplitude(imaginary unit $\bm{\dot{\iota}}$ in eq.(\ref{quaternion}). These {\bf real quaternionic scattering matrix elements}  has been extensively used in past for studying symmetry classification of random matrices\cite{Dyson-Mehta, Altland} and it application to charge transport phenomena in mesoscopic systems\cite{RM-Beena} and has been also used to study spin-transport phenomena during past decade\cite{Kirch}. 
Therefore demanding that coefficients ${\cal{Q}}^{\alpha\beta}_{0ijk}$ are zero is equivalent to choosing scattering matrix elements to be {\bf real quaternion} and is equivalent to having an effective {\bf single scattering phase} which is what we call {\bf abelian} hydrodynamics, while  ${\cal{Q}}^{\alpha\beta}_{0ijk}\neq 0$ is equivalent to having a $2 \times 2$ complex matrix where each terms has an independent phase which is what we call {\bf non-abelian} scattering phases. 
To see this more clearly let us look at the expression ${\cal{S}}^{*}_{\alpha\beta} {\cal{S}}_{\alpha\beta}$,
\begin{eqnarray}
{\cal{S}}^{*}_{\alpha\beta} {\cal{S}}_{\alpha\beta}&=&(S^{*}_{\alpha\beta,0} \bm{I}_{2}+ \bm{S}^{*}_{\alpha\beta} \cdot \bm{\sigma})(S_{\alpha\beta,0}\bm{I}_{2}+ \bm{S}_{\alpha\beta} \cdot \bm{\sigma}) \nonumber \\
&=& (|S_{\alpha\beta,0}|^{2}+\sum_{i}|S_{i,\alpha \beta}|^{2})\bm{I}_{2}^{s}+[(S_{\alpha\beta,0}^{*}\bm{S}_{\alpha\beta}+\bm{S}_{\alpha\beta}^{*}S_{\alpha\beta,0})+\bm{\dot{\iota}}(\bm{S}_{\alpha\beta}^{*} \times \bm{S}_{\alpha\beta})]\cdot \bm{\sigma} \nonumber \\
&=&{\cal{C}}^{\alpha\beta} \bm{I}_{2}^{s}+[\sum_{i} 2\text{Re}(S_{\alpha\beta,0}{S}_{\alpha\beta,i}^{*})+\bm{\dot{\iota}}\sum_{(ijk)_{c}}(S_{\alpha\beta,j}^{*}S_{\alpha\beta,k}-S_{\alpha\beta,k}^{*}S_{\alpha\beta,j})]\sigma_{i} \nonumber \\
&=&{\cal{C}}^{\alpha\beta} \bm{I}_{2}^{s}+\sum_{(ijk)_{c}}2[\text{Re}(S_{\alpha\beta,0}{S}_{\alpha\beta,i}^{*}+\text{Im}(S_{\alpha\beta,j}S_{\alpha\beta,k}^{*})]\sigma_{i} \label{eq:AB-NAB-3}
\end{eqnarray}
Adding and subtracting eq. (\ref{eq:AB-NAB-1}) and (\ref{eq:AB-NAB-3}) we obtain commutator and anti-commutator associated with the scattering matrix element ${\cal{S}}_{\alpha\beta}$,
\begin{equation}
[{\cal{S}}_{\alpha\beta}^{*},{\cal{S}}_{\alpha\beta}]_{+} = 2{\cal{C}}^{\alpha\beta} \bm{I}_{2}^{s}+ 2 \sum_{(ijk)_{c}} 2 [\text{Re}(S_{\alpha\beta,0}{S}_{\alpha\beta,i}^{*})] \sigma_{i}
\end{equation}
\begin{equation}
[{\cal{S}}_{\alpha\beta}^{*},{\cal{S}}_{\alpha\beta}]_{-}  =  2 \sum_{(ijk)_{c}} 2 [\text{Im} ({S}_{\alpha\beta,j}{S}_{\alpha\beta,k}^{*})]\sigma_{i}
\end{equation}
where $[--, --]_{-}$ stands for commutator and $[--, --]_{+}$ anti-commutator. Therefore the {\bf quantum non-abelian matrix}  ${\cal{Q}}^{\alpha\beta}_{0ijk}\sigma_{i}$'s can be written in terms of above commutator and  anti-commutator as,
\begin{eqnarray}
\frac{1}{4}\{[{\cal{S}}_{\alpha\beta}^{*},{\cal{S}}_{\alpha\beta}]_{+}+[{\cal{S}}_{\alpha\beta}^{*},{\cal{S}}_{\alpha\beta}]_{-}\}-\frac{1}{2}{\cal{C}}^{\alpha\beta} \bm{I}_{2}^{s}&=&\sum_{(ijk)_{c}}2{\cal{Q}}^{\alpha\beta}_{0ijk}\sigma_{i}. \\
%&\equiv& \left[\begin{array}{cc}
                     % {\cal{Q}}_{0zxy}^{\alpha\beta} & {\cal{Q}}_{0xyz}^{\alpha\beta}-\bm{\dot{\iota}}{\cal{Q}}_{0yzx}^{\alpha\beta}\\
                     % {\cal{Q}}_{0xyz}^{\alpha\beta}+\bm{\dot{\iota}}{\cal{Q}}_{0yzx}^{\alpha\beta}& -{\cal{Q}}_{0zxy}^{\alpha\beta}
                     % \end{array} \right]
\end{eqnarray}
Therefore non-zero values of {\bf quantum coefficients} are related to the fact that scattering matrix elements which themselves are matrices neither commutes nor anti-commutes. Therefore each scattering matrix element ${\cal{S}}_{\alpha\beta}$ is equivalent to a transformation with is neither commutative nor anti-commutative which is what is taken as defining property of the abelian or non-Abelian statistics see ref.() 
Since the scattering matrix elements governs the transition probabilities therefore the quasi particle excitations of a {\bf Quantum Non-Abelian} theory would be neither fermionic nor bosonic rather it will be {\bf Anyonic} in nature. This leads to {\bf charge fractionalization}, which will be further discussed using number or charge density conservation condition in section (\ref{GLCL}).

The various interrelated features discusses above such as {\bf Non-abelian phase}, {\bf Spin-orbital entanglement} and {\bf Anyonoic} nature of quasi particle  excitation can all be combined into a {\bf Vector Order Parameter}, which is defined using the expression (\ref{eq:DP-VQ1}) as,
\begin{eqnarray}
\bm{Q}=\sum_{\alpha\beta}\bm{Q}^{\alpha\beta}&=&\sum_{\alpha\beta,(ijk)_{c}}{\cal{Q}}^{\alpha\beta}_{0ijk}\hat{i} \nonumber \\
&\equiv& \sum_{\alpha\beta}[{\cal{Q}}^{\alpha\beta}_{0xyz}\hat{x}+{\cal{Q}}^{\alpha\beta}_{0yzx}\hat{y}+{\cal{Q}}^{\alpha\beta}_{0zxy}\hat{z}]
\end{eqnarray}
Now we proceed to show that $\bm{Q}\neq 0$ corresponds to {\bf Abelian} hydrodynamics and is equivalent to having a unitary scattering matrix while  $\bm{Q}\neq 0$ corresponds to {\bf Non-Unitarity} scattering matrix and {\bf Non-Abelian} hydrodynamics. Toward this end we note that the
hermitian matrices, ${\cal{SS}}^{\dagger}$ and ${\cal{S}}^{\dagger}\cal{S}$ which can be obtained straight forwardly
using the defining relation eq.(\ref{eq:SM-3}), are given by, 
\begin{eqnarray}
 {\cal{S}}{\cal{S}}^{\dagger}&=&\left[S_{0}S_{0}^{\dagger}+\bm{S}\cdot\bm{S}^{\dagger}\right]\bm{I}_{2}+\left[\{S_{0}\bm{S}^{\dagger}
+\bm{S}S_{0}^{\dagger}\}+\bm{\dot{\iota}} \{\bm{S}\times \bm{S}^{\dagger}\} \right]\cdot \bm{\sigma} \label{eq:tr-SM-1}\\
&=& A_{0}\otimes \bm{I}_{2}+\sum_{i}A_{i}\otimes \sigma_{i}\equiv  A_{0}\bm{I}_{2}+\bm{A}\cdot \bm{\sigma} 
\label{eq:tr-SM-2} \\
\cal{S}^{\dagger}\cal{S}&=&\left[S_{0}^{\dagger}S_{0}+\bm{S}^{\dagger}\cdot\bm{S}\right]\bm{I}_{2}+\left[\{\bm{S}^{\dagger}{S}_{0}
+S_{0}^{\dagger}\bm{S}\}+\bm{\dot{\iota}} \{\bm{S}^{\dagger}\times \bm{S}\} \right]\cdot \bm{\sigma} \label{eq:tr-SM-3} \\
&=& B_{0}\otimes \bm{I}_{2}+\sum_{i}B_{i}\otimes \sigma_{i}\equiv  B_{0}\bm{I}_{2}+\bm{B}\cdot \bm{\sigma} 
\label{eq:tr-SM-4}.
\end{eqnarray}
From the above expression we see that,  ${\cal{S}\cal{S}^{\dagger}}$ and  ${\cal{S}^{\dagger}\cal{S}}$ are composed of scalar matrices $(A_{0},B_{0})$ and vectorial matrices $(\bm{A},\bm{B})$ respectively, which is similar to the structure of incident density matrix (eq.(\ref{eq:IDM-3})), scattering matrix(eq.(\ref{eq:SM-3})) and scattered density matrix (eq.(\ref{eq:SDM-3})) discusses before. Moreover the matrices ${\cal{S}\cal{S}^{\dagger}}$ and  ${\cal{S}^{\dagger}\cal{S}}$ are hermitian. 
Now it is straight forward to see that the following trace relations holds (the details are given in appendix),
\begin{eqnarray}
\text{Tr}[{\cal{S}}{\cal{S}}^{\dagger}]=2\text{tr}[A_{0}]&=&\text{Tr}[{\cal{S}}^{\dagger}{\cal{S}}]=2\text{tr}[B_{0}] \nonumber\\
&=&2 \sum_{i,\alpha, \beta} \left[ (|S_{\alpha \beta,0}|^{2} +|S_{\alpha \beta,i}|^{2})\right] \equiv \sum_{\alpha\beta,i} {\cal{C}}_{0i}^{\alpha\beta} \label{eq:tr-SM-5}\\
\text{Tr}[{\cal{S}}{\cal{S}}^{\dagger}\bm{\sigma}]=2\text{tr}[\bm{A}]&=&2\text{tr}[\{S_{0}\bm{S}^{\dagger}
+\bm{S}S_{0}^{\dagger}\}+\bm{\dot{\iota}} \{\bm{S}\times \bm{S}^{\dagger}\}] \label{eq:tr-SM-6} \\
&=&2\sum_{\{ijk\},\alpha\beta}[2\text{Re}(S_{\alpha\beta,0}S_{\alpha\beta,i}^{*})-2 \text{Im}(S_{\alpha\beta,j}{S}_{\alpha\beta,k}^{*})]\hat{i} \label{eq:tr-SM-7}\\
&=&2\sum_{\{ijk\},\alpha\beta}[2{\cal{Q}}^{\alpha\beta}_{0ijk}]\hat{i}\label {eq:tr-SM-8} \\
\text{Tr}[{\cal{S}}^{\dagger}{\cal{S}}\bm{\sigma}]=2\text{tr}[\bm{B}]&=&2\text{tr}\left[\{\bm{S}^{\dagger}\bm{S}_{0}
+S_{0}^{\dagger}\bm{S}\}+\bm{\dot{\iota}} \{\bm{S}^{\dagger}\times \bm{S}\} \right] \label{eq:tr-SM-9} \\
&=&2\sum_{\{ijk\},\alpha\beta}[2\text{Re}(S_{\alpha\beta,0}S_{\alpha\beta,i}^{*})+2 \text{Im}(S_{\alpha\beta,j}{S}_{\alpha\beta,k}^{*})]\hat{i} \label{eq:tr-SM-10}. \\
&=&2\sum_{\{ijk\},\alpha\beta}[2{\cal{Q}}^{\alpha\beta,\text{rev}}_{0ijk}]\hat{i}\label {eq:tr-SM-11} 
\end{eqnarray}
First we note that ${\cal{Q}}^{\alpha\beta}_{0ijk} \propto \text{Tr}[{\cal{S}}{\cal{S}}^{\dagger}\bm{\sigma}]$ which occurs in charge density expression (\ref{eq:DP-1}). On the other hand $\text{Tr}[{\cal{S}}^{\dagger}{\cal{S}}\bm{\sigma}]\propto {\cal{Q}}^{\alpha\beta,\text{rev}}_{0ijk} \neq {\cal{Q}}^{\alpha\beta}_{0ijk} $, these coefficients occurs in the trace expression for the time reversed scattered density matrix as we will see in section(\ref{sec:time-rev}). The important point to note is that {\bf time reversal symmetry is broken} for scattered density matrix which will shown explicitly in section(\ref{sec:time-rev}). In-fact we will show that if time reversal symmetry is imposed on the scattered density matrix 
it requires that ${\cal{Q}}^{\alpha\beta,\text{rev}}_{0ijk} ={\cal{Q}}^{\alpha\beta}_{0ijk}$ which would require that $\text{Im}(S_{\alpha\beta,j}{S}_{\alpha\beta,k}^{*})=0$ which is only possible for {\bf Abelian} hydrodynamics as was discussed previously.
 
Though the quantum coefficients  ${\cal{Q}}^{\alpha\beta}_{0ijk}$ and ${\cal{Q}}^{\alpha\beta,\text{rev}}_{0ijk}$ are not identical,however, they are composed of same terms. Therefore if ${\cal{Q}}^{\alpha\beta}_{0ijk}$ will be zero if vectorial matrices $\bm{A}$ and $\bm{B}$ are traceless. Note that this does not imply that scattering matrix is unitary which requires that vectorial matrices $\bm{A}$ and $\bm{B}$ are identically zero, i.e., $\bm{A}=\bm{B}\equiv 0$, and, scalar matrices  $A_{0}$ and $B_{0}$ are equal to identity matrix,i.e,  $A_{0}=B_{0}\equiv {\cal{I}}$. This requires that following matrix constraints among scalar and components of vectorial scattering matrix should be satisfied,i.e,
\begin{eqnarray}
A_{0}=B_{0}&\equiv&   S_{0}S_{0}^{\dagger}+\sum_{i}{S}_{i}{S}_{i}^{\dagger} = S_{0}^{\dagger}S_{0}+\sum_{i}{S}_{i}^{\dagger}{S}_{i} \nonumber \\
&\implies& \left[S_{0},S_{0}^{\dagger}\right]_{-}=\sum_{i}\left[{S}_{i}^{\dagger},{S}_{i}\right]_{-} \label{eq:A0=B0}\\
\bm{A}=\bm{B} &\equiv & A_{i} = B_{i} \nonumber \\
&\implies& S_{0}S_{i}^{\dagger}+S_{i}S_{0}^{\dagger}+\bm{\dot{\iota}}(S_{j}S_{k}^{\dagger}-S_{k}S_{j}^{\dagger})=S_{i}^{\dagger}S_{0}+S_{0}^{\dagger}S_{i}+\bm{\dot{\iota}}(S_{j}^{\dagger}S_{k}-S_{k}^{\dagger}S_{j}) \nonumber \\
&\equiv& \left[S_{0},S_{i}^{\dagger}\right]_{-}+\left[S_{i}, S_{0}^{\dagger}\right]_{-}=\bm{\dot{\iota}}\left\{\left[S_{j}^{\dagger},S_{k}\right]_{+}-\left[S_{k}^{\dagger}, S_{j}\right]_{+}\right\}.\label{eq:A=B}
\end{eqnarray}
Therefore unitarity condition is much stronger and rather unphysical condition as it requires that commutator and anti-commutator of matrices, $S_{0}$, $S_{0}^{\dagger}$ and $S_{i}$ $S_{i}^{\dagger}$
(i =x,y,z) should satisfy relation given in eq.(\ref{eq:A0=B0})and eq.({eq:A=B}) which is much stronger than the requirement that scattering matrix elements are real quaternionic as was discussed above.

The previous discussion can be summarized in a following equivalent way in terms of {\bf Vector Order Parameter}, $\bm{Q}$. 
Adding and subtracting eqs.(\ref{eq:tr-SM-6}) and (\ref{eq:tr-SM-8}) leads to, 
\begin{eqnarray}
\text{Tr}[({\cal{S}}{\cal{S}}^{\dagger}+{\cal{S}}^{\dagger}{\cal{S}})\bm{\sigma}]&=&2\sum_{\{ijk\},\alpha\beta}[2\text{Re}(S_{0,\alpha\beta}S_{i,\alpha\beta}^{*})]\hat{i}  \label{eq:tr-SM-10-1} \\
\text{Tr}[({\cal{S}}^{\dagger}{\cal{S}}-{\cal{S}}{\cal{S}}^{\dagger})\bm{\sigma}]&=&2\sum_{\{ijk\},\alpha\beta}[2 \text{Im}(S_{j,\alpha\beta}{S}_{k,\alpha\beta}^{*})]\hat{i}\label{eq:tr-SM-11-1}.
\end{eqnarray}
Therefore the {\bf Vector Order Parameter}, $\bm{Q}$ can be expressed as, 
\begin{eqnarray}
2\bm{Q}&=& \text{Tr}[({\cal{S}}{\cal{S}}^{\dagger}+{\cal{S}}^{\dagger}{\cal{S}})\bm{\sigma}]+\text{Tr}[({\cal{S}}^{\dagger}{\cal{S}}-{\cal{S}}{\cal{S}}^{\dagger})\bm{\sigma}]\nonumber \\
&=&\text{Tr}\left[[{\cal{S}}^{\dagger},{\cal{S}}]_{+} \bm{\sigma}\right]+\text{Tr}\left[[{\cal{S}}^{\dagger},{\cal{S}}]_{-} \bm{\sigma}\right]
\label{Vec-Q}.
\end{eqnarray}
Now it is straight forward to see using eqs.(\ref{eq:tr-SM-10}-\ref{Vec-Q}) that if  unitarity of scattering matrix is assumed(rather forced),i.e, ${\cal{S}}{\cal{S}}^{\dagger}={\cal{S}}^{\dagger}{\cal{S}}=\cal{I}$(Identity matrix), it implies following three condition,
\begin{equation}
{\cal{S}}{\cal{S}}^{\dagger}={\cal{S}}^{\dagger}{\cal{S}}=\cal{I} \implies 
\begin{cases}  \left[{\cal{S}},{\cal{S}}^{\dagger}\right]_{+}=2 \cal{I} \\
\left[{\cal{S}},{\cal{S}}^{\dagger}\right]_{-}=0 \\
\bm{Q}=0
\end{cases}\label{Vec-Q-1}
\end{equation}
where the first condition says that anti-commutator of scattering matrix ${\cal{S}}$ with its dagger ${\cal{S}}^{\dagger}$ is unity while its commutator is zero(the second condition above) and these two conditions are equivalent to having $\bm{Q}=0$,i.e, {\bf vector order parameter} vanishes. 

The other case of {\bf Non-Unitary} scattering matrix implies that ${\cal{S}}{\cal{S}}^{\dagger}\neq {\cal{S}}^{\dagger}{\cal{S}}\neq \cal{I}$ as can be seen from corresponding expressions (\ref{eq:tr-SM-2}-\ref{eq:tr-SM-4}) which corresponds to following three conditions, 
\begin{eqnarray}
{\cal{S}}{\cal{S}}^{\dagger}\neq{\cal{S}}^{\dagger}{\cal{S}}\neq \cal{I} \implies 
\begin{cases}
\left[{\cal{S}},{\cal{S}}^{\dagger}\right]_{+}=[A_{0}+B_{0}]\bm{I}_{2}^{s}+[\bm{A}+\bm{B}]\cdot \bm{\sigma}\\
\left[{\cal{S}},{\cal{S}}^{\dagger}\right]_{-}=[A_{0}-B_{0}]\bm{I}_{2}^{s}+[\bm{A}-\bm{B}]\cdot \bm{\sigma} \\
\bm{Q}\neq 0
\end{cases}\label{Vec-Q-2}
\end{eqnarray}
Therefore a non zero values of vector order parameter $\bm{Q}$ corresponds to having both commutator and anti commutator of scattering matrix ${\cal{S}}$ with its dagger ${\cal{S}}^{\dagger}$ to be non-zero as well different from identity matrix which is only possible if scattering phases are {\bf Non-Abelian} as was discussed in the beginning of this section. Therefore $\bm{Q}\neq 0$ implies that quasi particle excitations will be Anyonic in nature. A pictorial representation of above two conditions[eqs.(\ref{Vec-Q-1}-\ref{Vec-Q-2})] on vector order parameter $\bm{Q}$ and physical phenomena associated with it
is provided in Fig.(3).
\begin{center}
\begin{figure}
\includegraphics[width=\linewidth,height=3.0in,angle=0,keepaspectratio]{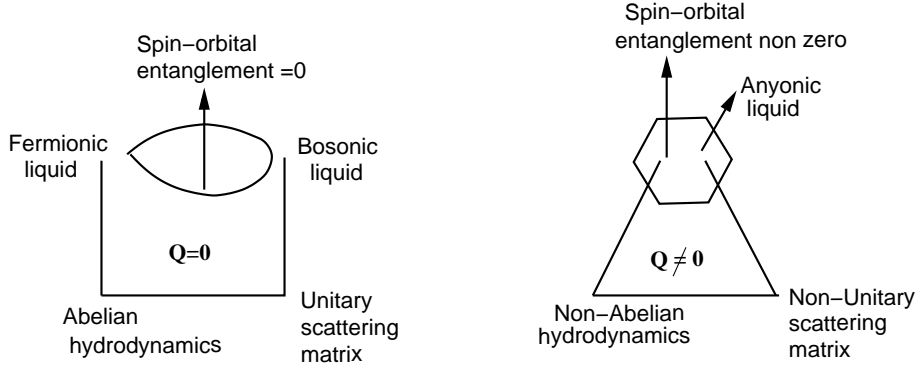}
\caption{\label{fig2} The pictorial representation of Vectorial Order Parameter $\bm{Q}$. {\it Left Panel}: $\bm{Q}=0$ corresponds to {\bf Abelian and Unitary} hydrodynamics which is equivalent to absence of spin-orbital entanglement consequently the fundamental excitations are Fermionic or Bosonic. {\it Right Panel} $\bm{Q}\neq 0$  corresponds to {\bf Non-Abelian and Non-Unitary} hydrodynamics and is equivalent to having a non-zero spin-orbital entanglement because of this the liquid is {\bf Anyonic}}
\end{figure}
\end{center}

In the section below we look at 
these phenomena from the charge density conservation perspective which provides another view point and strength the previous discussion and also gives a clear picture about charge fractionalization.

\section{The Charge Fractionalization via Charge density conservation}
\label{GLCL}
The seemingly self contradictory title of this section will be best appreciated once we look at the charge conservation constraints which requires that the trace of incident density matrix and scattered density matrix should be same,
i.e, 
\begin{eqnarray}
\text{Tr}(\rho_{in})=\text{Tr}( \rho_{sc})\implies \text{tr}(\rho_{in})&=&\text{tr}(\varrho_{sc,0})\\
\sum_{\beta}\rho_{in,\beta,0}=\rho_{in,L,0}+\rho_{in,R,0}&=& \sum_{\alpha\beta}\left[{\cal{C}}^{\alpha\beta} \rho_{in,\beta,0}  \pm 2 |\bm{Q}^{\alpha\beta}\cdot \bm{p}_{in,\beta}|\right] \\
&=&\sum_{\alpha\beta}\left[{\cal{C}}^{\alpha\beta} \rho_{in,\beta,0}  \pm 2 |\bm{Q}^{\alpha\beta}\cdot {\hat{\bm{n}}}_{\beta}| {p}_{in,\beta}\right] \label{eq:glcl-1}
\end{eqnarray}
where we have made use of eqs.(\ref{eq:IDM-4}) and (\ref{eq:DP-1.2})[equivalent expression(\ref{eq:DP-1})] and by expression incident polarization vector as, $\bm{p}_{in,\beta}= {\hat{\bm{n}}}_{\beta} {p}_{in,\beta}$, where $0 < {p}_{in,\beta} < 0$ is the magnitude of incident polarization and ${\hat{\bm{n}}}_{\beta}$ is quantization axis we arrive at the last expression. As discussed previously for the case of incident fluid being un-polarized is not equivalent to putting ${p}_{in,\beta}=0$ for the scattered charge density.
The quantity $|\bm{Q}^{\alpha\beta}\cdot \hat{\bm{n}}_{in,\beta}|$ is modulus of the real number $\bm{Q}^{\alpha\beta}\cdot \hat{\bm{n}}_{in,\beta} $ which can positive as well negative while $ \sum_{\alpha}{\cal{C}}^{\alpha\beta} > 0$ are positive real numbers. 
Physically the above condition implies that charge or number density is conserved globally. 
The local version of charge density conservation requires that for a particular lead (left or right), i.e, for fixed $\beta$ the following condition should be satisfied,
\begin{eqnarray}
\rho_{in,\beta,0}&=& [\sum_{\alpha}{\cal{C}}^{\alpha\beta}] \rho_{in,\beta,0}  \pm \sum_{\alpha}2  | \bm{Q}^{\alpha\beta}\cdot {\hat{\bm{n}}}_{\beta}| {p}_{in,\beta}   
 \label{eq:glcl-2}
\end{eqnarray}
where, $\alpha \in \{L, R\}$ while $\beta$ takes values either, $L$, or $R$. 
The above equation gives the charge density in lead $\beta$ after the scattering or for the scattered fluid. Now since  $\bm{Q}^{\alpha\beta}\neq 0$ and $\rho_{in,\beta,0}$ has to be a positive real number therefore the solution of above equation for $\rho_{in,\beta,0}$  are,
\begin{eqnarray}
\rho_{in,\beta,0}= \frac{\sum_{\alpha}2|{\bm{Q}}^{\alpha\beta}\cdot \hat{\bm{n}}_{\beta}|{p}_{in,\beta}}{1-[\sum_{\alpha}{\cal{C}}^{\alpha\beta}]} \,\,\,\,\,\text{for}\,\,  \sum_{\alpha}{\cal{C}}^{\alpha\beta} < 1, \text{and},\,  ({\bm{Q}}^{\alpha\beta} \cdot \bm{p}_{in,\beta}) > 0 
\end{eqnarray}
\begin{equation}
\rho_{in,\beta,0}=\frac{\sum_{\alpha}2|{\bm{Q}}^{\alpha\beta}\cdot \hat{\bm{n}}_{\beta}|{p}_{in,\beta}}{[\sum_{\alpha}{\cal{C}}^{\alpha\beta}]-1}, \,\,\, \,\,\text{for}\,\,  \sum_{\alpha}{\cal{C}}^{\alpha\beta} > 1, \text{and},\,  ({\bm{Q}}^{\alpha\beta} \cdot \bm{p}_{in,\beta}) < 0 \\
\end{equation}  
Therefore if the incident fluid is polarized the charge density need not be an integer. 

Now if the lead $\beta$ is connected to non-magnetic reservoir which implies the incident fluid is un-polarized in the lead, for this case one should take ${p}_{in,\beta}=1$ and integrate over all possible initial quantization axis as was discussed in the last paragraph of section (\ref{sec-Num-DP}) from physical consideration and in (\ref{sec-SU2-inv}) from broken SU(2) symmetry point of view. Therefore the solution for $\rho_{in,\beta,0}$  when incident fluid is un-polarized is given by,
\begin{equation}
\rho_{in,\beta,0}= \frac{ \sum_{\alpha} 2|\int ({\bm{Q}}^{\alpha\beta}\cdot \hat{\bm{n}}_{\beta})d\hat{\bm{n}}_{\beta}|}{1-\sum_{\alpha}[\int{\cal{C}}^{\alpha\beta}d\hat{\bm{n}}_{\beta}]]} \,\,\,\,\,\text{for}\,\,  \int\sum_{\alpha}{\cal{C}}^{\alpha\beta}d\hat{\bm{n}}_{\beta} < 1, \text{and},\,  \int({\bm{Q}}^{\alpha\beta} \cdot \hat{\bm{n}}_{in,\beta})d\hat{\bm{n}}_{\beta} > 0  
\end{equation}
\begin{equation}
\rho_{in,\beta,0}= \frac{ \sum_{\alpha} 2|\int ({\bm{Q}}^{\alpha\beta}\cdot \hat{\bm{n}}_{\beta})d\hat{\bm{n}}_{\beta}|} {\sum_{\alpha}[\int{\cal{C}}^{\alpha\beta}d\hat{\bm{n}}_{\beta}]-1} \,\,\,\,\,\text{for}\,\,  \int\sum_{\alpha}{\cal{C}}^{\alpha\beta}d\hat{\bm{n}}_{\beta} > 1, \text{and},\, \int({\bm{Q}}^{\alpha\beta} \cdot \hat{\bm{n}}_{\beta})d\hat{\bm{n}}_{\beta} < 0. 
\end{equation}
Hence we have show that charge fractionalization is a natural consequence of ${\bm{Q}}^{\alpha\beta}\neq 0$ which is equivalent to, scattering matrix being {\bf non-unitary}, scattering phases being {\bf non-abelian} and {\bf spin-space entanglement} being non-zero. In-fact the  conductance measurement in low dimensional system amounts to projecting the scalar part of scattered density matrix $\rho_{sc}$(in eq.(\ref{eq:SDM-2}) on a specific lead. However since $\rho_{sc}$ is spin-space entangled, therefore, it is
not separable even in lead indices, hence, the projection on a particular lead equivalently reservoir will be sensitive to local charge or number density of the ``Entangled Bloch Sphere``  which reflects as the fractional conductance in the experiments \cite{Cronenwett,Danneau,Hew,Quay,Debray,Wan}.
the  will only disturb the full $\rho_{sc,0}$ which is what leads to conductance fractionalization seen in the low dimensional system\cite{}.

In presence of spin-dependent scattering it is expected on physical grounds that spin-density vector is not conserved,i.e,
\begin{equation}
\bm{P}_{in}=\text{Tr}(\rho^{in}\bm{\sigma})\neq \bm{P}_{sc}=\text{Tr}( \rho_{sc}\bm{\sigma}),  \label{eq:SD-C-NC1}
\end{equation}
which is also explicitly obvious from the expression (\ref{eq:DP-SD}) for spin-density vector in scattered fluid. Therefore we do need to consider spin-density conservation or non-conservation separately.

Before we end this section, for sake of completeness we note that when scattering is spin-independent(potential being spin-independent) and the scattering matrix reduced to a scalar matrix, i.e., $\bm{S}\equiv 0$ in eq.(\ref{eq:SM-3}), equivalently $S_{i}\equiv 0 \, \forall \, i\, \in (x,y,z)$ which implies that and all quantum interference coefficients ${\cal{Q}}$'s appearing in charge density and spin density expression are zero. Therefore the local charge density
eq.(\ref{eq:glcl-2}) and the spin density of scattered fluid given in eq.(\ref{eq:DP-SD}) reduces to,
\begin{eqnarray}
\rho_{in,\beta,0}&=&[\sum_{\alpha}{\cal{C}}^{\alpha\beta}] \rho_{in,\beta,0} \\
\text{tr}(\bm{\varrho}_{sc})&=& \sum_{\beta}[\sum_{\alpha}{\cal{C}}^{\alpha\beta}]\bm{p}_{in,\beta}\equiv \sum_{\beta}\bm{P}_{sc,\beta}. 
\end{eqnarray}
Now if local charge density is to be conserved it requires, $\sum_{\alpha}{\cal{C}}^{\alpha\beta}=1$ which is nothing but unitarity of scattering matrix in absence of spin-dependent interaction. We note that local charge density conservation also implies the global charge density conservation
and vice versa, this is true only spin-independent case, as can be seen from the eq.(\ref{eq:glcl-1}).
Moreover this also implies that 
in absence of spin-dependent interaction spin-density vector is separately conserved in each lead both in magnitude as well in direction as is seen from the second expression above. Physically this implies that polarization vector in each lead remains static,i.e, scattering does not introduce any rotation of polarization vector as well its magnitude also remains constant. 
We note importantly that requiring $\sum_{\alpha}{\cal{C}}^{\alpha\beta}=1$ only implies that diagonal elements of scattered density matrix are 
unity, however, it does not imply that off-diagonal elements are zero. Infact for the spin-independent case having off-diagonal elements non-zero 
implies that left and right spatial region are entangled. However this spatial entanglement is nothing but the spatial phase coherence which gets defined by the spatial phase coherence length and gives rise to AB oscillation of ring shaped conductors. Of course this aspect can be discussed in 
purely academic terms which will be done in section.(\ref{sec:GEL})

In the next section we consider time reversal operation and look at the scattered density matrix for the time reversed situation which shows explicitly that time reversal symmetry spin-density vector and discusses its conservation and non-conservation which provides a physical picture 
for the discussions and mathematical analysis present in various sections before.

\section{Time Reversal Operation} 
\label{sec:time-rev}
Since the whole previous discussion is based on the density matrix - incident and scattered. Therefore it is natural that we define time reversal operation with respect to density matrix. 
To do this let us look at the incident density matrix  given in two equivalent forms in eqs.(\ref{eq:IDM-2}) and (\ref{eq:IDM-3}). Physically it is expected that under the action of time reversal the charge(number) density should remain invariant while spin-density vector should reverse or change sign. The is achieved using time reversal operator, ${\cal{T}}={\cal{Z}}{\cal{K}}\equiv \bm{I}_{lead}\otimes (-\bm{\dot{\iota}}\sigma_{y}){\cal{K}}$, where $\bm{I}_{lead}$ is identity operator in lead space and $ {\cal{K}}$ is complex conjugation operator. The application of time reversal operator on incident density matrix eq.(\ref{eq:IDM-3}) leads to,
\begin{eqnarray}
{\widetilde{\varrho_{in}}}&=& {\cal{T}}\varrho_{in}(\hat{\bm{n}}_{\alpha},\hat{\bm{n}}_{\beta}){\cal{T}}^{-1}=(1/2){\cal{Z}}{\cal{K}}\rho_{in}(\hat{\bm{n}}_{\alpha},\hat{\bm{n}}_{\beta}){\cal{K}}^{-1}{\cal{Z}}^{-1}\nonumber \\
&=&(1/2){\cal{Z}}[\rho_{in}\otimes\bm{I}^{s}_{2}+\sum_{k \in\{x,y,z\}}{\varrho}_{k}\otimes{\sigma}_{k}^{*}]{\cal{Z}} \nonumber \\
&=&(1/2)[\rho_{in}\otimes\bm{I}^{s}_{2}-\sum_{k \in\{x,y,z\}}{\varrho}_{k}\otimes{\sigma}_{k}]
\equiv (1/2)[\rho_{in}\bm{I}^{s}_{2}-{\bm{\rho}}_{in}\cdot \bm{\sigma}]
\label{eq:time-rev-1}
\end{eqnarray}
where, ${\widetilde{\varrho_{in}}}$, is the time reversed incident density matrix. Now it is straight forward to see that under time reversal operation the charge(number) density remains invariant while spin-density vector changes sign, i.e,
\begin{eqnarray}
\text{Tr}({\widetilde{\varrho_{in}}})=\text{Tr}(\varrho_{in})=\rho_{in,L,0}+\rho_{in,R,0} \\
\text{Tr}({\widetilde{\varrho_{in}}}\bm{\sigma})=-\text{tr}({\bm{\rho}}_{in})=-\text{Tr}(\varrho_{in}\bm{\sigma})=-(\bm{p}_{in,L}+\bm{p}_{in,R})
\end{eqnarray}
in obtaining above relation we have made use of eqs.(\ref{eq:IDM-4}) and (\ref{eq:IDM-5}). But it is important to realize that time reversed density matrix corresponds to different physical situation, because for the incident density matrix the spin-density vector points in opposite direction.

Therefore we have defined a physically appropriate time reversal operator, i.e,${\cal{T}}={\cal{Z}}{\cal{K}}\equiv \bm{I}_{lead}\otimes (-\bm{\dot{\iota}}\sigma_{y}){\cal{K}}$. 
Now we apply this time reversal operation on scattering matrix and scattered density matrix.
The scattering matrix defined in eq.(\ref{eq:SM-3}) and it conjugate transforms under time reversal operation as,
\begin{eqnarray}
{\widetilde{\cal{S}}}={\cal{T}}{\cal{S}}{\cal{T}}^{-1}=S_{0}^{*}\bm{I}_{2}^{s}-\bm{S}^{*}\cdot \bm{\sigma} \\
{\widetilde{\cal{S}}^{\dagger}}={\cal{T}}{\cal{S}^{\dagger}}{\cal{T}}^{-1}=S_{0}^{T}\bm{I}_{2}^{s}-\bm{S}^{T}\cdot \bm{\sigma} 
\end{eqnarray}
where ,$*$, and $T$, represents complex conjugation and transpose operation. 
Therefore the time reversed scattered density matrix is given by,
\begin{eqnarray}
{\widetilde{\varrho_{sc}}}&=&{\cal{T}}[{\cal{S}}\varrho_{in}{\cal{S}^{\dagger}}]{\cal{T}}^{-1}\equiv {\widetilde{\cal{S}}}{\widetilde{\varrho_{in}}}{\widetilde{\cal{S}}^{\dagger}} \nonumber \\
&=&[S_{0}^{*}\bm{I}_{2}^{s}-\bm{S}^{*} \cdot \bm{\sigma}] [\rho_{in}\bm{I}^{s}_{2}-{\bm{\rho}}_{in}\cdot \bm{\sigma}][S_{0}^{T}\bm{I}_{2}^{s}-\bm{S}^{T}\cdot \bm{\sigma}].\label{eq:SDM-time-rev}
\end{eqnarray}
We notice the above expression which corresponds to time reversed scattered density matrix is not same as that with the expression (\ref{eq:SDM-1}) obtained earlier, reflecting that time reversal symmetry is broken for the scattered density matrix. 
The simplified expression corresponding to eq.(\ref{eq:SDM-time-rev}) can be obtain if we replace $(S_{0},\bm{S})\mapsto (S_{0}^{*},-\bm{S}^{*})$,  
$(S_{0}^{\dagger},\bm{S}^{\dagger})\mapsto (S_{0}^{T},-\bm{S}^{T})$ and $(\rho_{in},\bm{\rho}_{in}) \mapsto (\rho_{in},-\bm{\rho}_{in})$ in the original expression given in eq.(\ref{eq:SDM-2}) which leads to,
\begin{eqnarray}
{\widetilde{\varrho_{sc}}}\hspace{-0.3cm}&=&\hspace{-0.3cm} [S_{0}^{*}\rho_{in}S_{0}^{T}+\bm{S}^{*}\rho_{in}\cdot\bm{S}^{T}+\{(\bm{S}^{*}\cdot{\bm{\rho}}_{in})S_{0}^{T}+
S_{0}^{*}({\bm{\rho}}_{in}\cdot\bm{S}^{T})\}
+\bm{\dot{\iota}}(\bm{S}^{*}\times{\bm{\rho}}_{in})\cdot \bm{S}^{T}]\otimes {I}_{2} \nonumber \\
\hspace{-0.3cm}&+&\hspace{-0.3cm}[-S_{0}^{*}{\bm{\rho}}_{in}S_{0}^{T}-\{S_{0}^{*}\rho_{in}\bm{S}^{T}
+\bm{S}^{*}\rho_{in}S_{0}^{T}\}+\{-(\bm{S}^{*}\cdot{\bm{\rho}}_{in})\bm{S}^{T}
+(\bm{S}^{*}\times {\bm{\rho}}_{in})\times \bm{S}^{T})\}]\cdot \bm{\sigma} \nonumber \\
\hspace{-0.3cm}&+&\hspace{-0.3cm}[\{\bm{\dot{\iota}}S_{0}^{*}({\bm{\rho}}_{in}\times \bm{S}^{T})+\bm{\dot{\iota}}(\bm{S}^{*}\times{\bm{\rho}}_{in})S_{0}^{T}\} 
+\bm{\dot{\iota}}(\bm{S}^{*}\rho_{in}\times \bm{S}^{T}) ]\cdot\bm{\sigma}. \\
\hspace{-0.3cm}&=&\hspace{-0.3cm} {\widetilde{\rho_{sc}}}\otimes {I}^{s}_{2}+\sum_{i}{\widetilde{{\rho}_{sc,i}}}\otimes{\sigma}_{i} \hat{i}\equiv {\widetilde{\rho_{sc}}}\otimes {I}^{s}_{2}+{\widetilde{\bm{\rho}_{sc}}}\cdot \bm{\sigma}
\label{time-rev-SDM}
\end{eqnarray}
Again the above expression is different from the original expression given in eq.(\ref{eq:SDM-2}) showing explicitly that time reversal symmetry is broken for the scattered density matrix. 

Now let us look at the trace for the time reversed scattered density matrix, toward we notice that scalar part of above expression (\ref{time-rev-SDM}),i.e, ${\widetilde{\rho_{sc}}}$ satisfies the following relation,
\begin{equation}
({\widetilde{\rho_{sc}}})^{*}=[S_{0}\rho_{in}S_{0}^{\dagger}+\bm{S}\rho_{in}\cdot\bm{S}^{\dagger}+\{(\bm{S}\cdot{\bm{\rho}}_{in})S_{0}^{\dagger}+
S_{0}({\bm{\rho}}_{in}\cdot\bm{S}^{\dagger})\}
-\bm{\dot{\iota}}(\bm{S}\times{\bm{\rho}}_{in})\cdot \bm{S}^{\dagger}]. 
\end{equation}
The above expression differs from the corresponding original expression in eq.(\ref{eq:SDM-2}) by a negative sign in the last term. Since ${\widetilde{\rho_{sc}}}$ is hermitian therefore,$\text{Tr}[{\widetilde{\rho_{sc}}}]=\text{Tr}[({\widetilde{\rho_{sc}}})^{*}]$,therefore the trace can be calculated straight forwardly and is given by,
\begin{eqnarray}
\text{Tr}[{\widetilde{\rho_{sc}}}]&= & 2 \sum_{\alpha\beta}\left[{\cal{C}}^{\alpha\beta} \rho_{in,\beta,0}  + \sum_{(ijk)_{c}}2{\cal{Q}}^{\alpha\beta,\text{rev}}_{0ijk}p_{in,\beta,i}\right] \\
{\cal{Q}}^{\alpha\beta,\text{rev}}_{0ijk}&=&[\text{Re}(S_{\alpha \beta,0}S_{\alpha \beta,i}^{*})+
\text{Im}(S_{\alpha\beta,j}^{*} S_{\alpha\beta,k})]\propto \text{Tr}[{\cal{S}}^{\dagger}{\cal{S}}\bm{\sigma}],\label{eq:DP-Q1-rev}. 
\end{eqnarray}
By comparing the above expression with eq.(\ref{eq:DP-1}) and (\ref{eq:DP-Q1}) we see that it only differs in terms of the {\it quantum coefficient}. In the original expression {\it quantum coefficient} ${\cal{Q}}^{\alpha\beta}_{0ijk}$ occurs which are related to the $\text{Tr}[{\cal{S}}{\cal{S}}^{\dagger}\bm{\sigma}]$ see , while for the time reversed scattered density matrix coefficients ${\cal{Q}}^{\alpha\beta,\text{rev}}_{0ijk}$ are related to the trace, $\text{Tr}[{\cal{S}}^{\dagger}{\cal{S}}\bm{\sigma}]$.

Now if one demands time reversal symmetry for the scattered density matrix which requires the trace expression (\ref{eq:DP-Q1-rev}) should be equal to trace expression given in the eq.(\ref{eq:DP-1}), which is only possible iff 
${\cal{Q}}^{\alpha\beta,\text{rev}}_{0ijk}={\cal{Q}}^{\alpha\beta,}_{0ijk}$, which requires that,
\begin{eqnarray}
\text{Im}(S_{\alpha\beta,j}^{*} S_{\alpha\beta,k}) = |S_{\alpha\beta,j}| |S_{\alpha\beta,k}| [\sin(\xi_{\alpha\beta,j})-\sin (\xi_{\alpha\beta,k})]\equiv 0.
\end{eqnarray}
The above condition for $|S_{\alpha\beta,j}|\neq 0$ and $ |S_{\alpha\beta,k})|\neq 0$ can only be satisfied, iff $\xi_{\alpha\beta,j}=\xi_{\alpha\beta,j}$ which is nothing but equivalent to {\bf abelian} hydrodynamics. Therefore we have shown that {\bf Non-Abelian} hydrodynamics necessary breaks the time reversal symmetry.

\section{Generic Entangled Liquids: Spin-Independent Case}
\label{sec:GEL}
Since ''entanglement'' has played a crucial role in the whole article, therefore, it should be possible to see the effect of entanglement even if the potential and the corresponding scattering matrix is spin-independent and when the incident or initial state of liquid is un-polarized. These three condition respectively corresponds to, $\bm{V}=0$ in eq.(\ref{eq:SP-1}), $\bm{S}=0$ in eq.(\ref{eq:SM-2}-\ref{eq:SM-3}) and $\bm{\rho}_{in}=0$ in eq.(\ref{eq:IDM-3}). Substituting these values  in the scattered density matrix expression eq.(\ref{eq:IDM-3}), one obtains the following simple expression,
\begin{equation}
\varrho_{sc}=[S_{0}\rho_{in}S_{0}^{\dagger}] 
\end{equation}
where $\rho_{in}$ is $2 \times 2$ scalar diagonal matrix[see eq.(\ref{eq:IDM-2})] and $S_{0}$ is $2 \times 2$ matrix defined in eq.(\ref{eq:SM-3}). Physically above density matrix corresponds to a two potential wells separated by a spin-independent potential step, see Fig.(2). These potential wells have an initial densities given by, $\rho_{in,L,0}$ and $\rho_{in,R,0}$, which are diagonal elements of the matrix 
${\rho}_{in}$ occurring in above expression. Now if we take  $\rho_{in,L,0}=1$ and $\rho_{in,R,0}=1$, which implies that initial densities in the wells are of unit magnitude. Then ${\rho}_{in}=\bm{I}$, where $\bm{I}$ is just an identity matrix in the real space. This reduces the density matrix to 
\begin{equation}
\varrho_{sc}=[S_{0}S_{0}^{\dagger}]  
\end{equation}
Therefore if the unitarity of scattering matrix$S_{0}S_{0}^{\dagger}=S_{0}^{\dagger}S_{0}=\bm{I}$  is assumed it will make the scattered density matrix to be a unit diagonal matrix. However scattered density matrix only needs to be hermitian , that would allow for the off-diagonal terms also in the matrix $S_{0}S_{0}^{\dagger}$  which is nothing but the density matrix. But this would necessarily imply that $S_{0}^{\dagger}S_{0}\neq S_{0}S_{0}^{\dagger}$. These simple arguments show that scattering matrix need not be unitary which is related to the fact that entanglement is non-zero. Here the ``entanglement`` is between left and right region, which are separated by a spin-independent potential step while in the previously discussed case ``entanglement`` was more general as it was between different degree of freedom of,i.e, spatial and spin as well between spatially separated region.
Therefore the analysis presented in previous section can be straight away carried forward for this case as well and will lead to the conclusion that scattering phases are non-abelian or equivalently the dynamics has to non-abelian.

In summary we can say that as long as {\bf Entanglement} is non-zero whether, the entanglement is between different parts of a system which are spatially separated as is the case in this section or between different parts as well between degrees of freedom of a system (spatial and spin degrees), it would imply that scattering matrix will be  non-unitary and scattering phase will be non-abelian.

\section{Conclusion}
In the present article we have studied non-adiabatic evolution of density matrix in presence of generic spin-dependent potential which is equivalent to an exact or non-perturbative scattering theory for density matrix. It is found that even if the initial density matrix is of product form(un-entangled) in the spatial and spin space, i.e,$\rho_{in}(\bm{r},\bm{\sigma})=\rho_{in}(\bm{r})\otimes \rho_{in}(\bm{\sigma})$, the final or scattered density matrix, $\rho_{sc}$, is no longer can be written as a product form, i.e., $\rho_{sc}(\bm{r},\bm{\sigma})\neq\rho_{sc}(\bm{r})\otimes \rho_{sc}(\bm{\sigma})$. Further it is shown that the $\rho_{sc}$ can be decomposed or expressed in electronic space as,
\begin{equation}
\rho_{sc}(\bm{r},\bm{\sigma})= \rho_{sc}(\bm{r},\bm{\sigma})\bm{I}_{2}^{s}+\bm{\rho}_{sc}(\bm{r},\bm{\sigma})\cdot \bm{\sigma}
\end{equation}
where the scalar part $\rho_{sc,0}$ and vectorial part $\bm{\rho}_{sc}$ are functions of both spatial degrees of freedom($\bm{r}$) as well internal
degree of freedom ($\bm{\sigma}$). 
In the present article we have used mesoscopic language to formulate the scattering theory. Specifically we have consider a two lead system connected to a spin-dependent scattering region which is equivalent to having two spin-independent potential wells separated by a spin-dependent matrix potential step. The spatial degree in our formulation has appeared in the form of lead indices,i.e, $\alpha$ and $\beta$. The non separability of $\bm{\rho}_{sc}$ with respect to lead indices as well spin-degrees makes it a $2 \times 2$ matrix. This is so because we have only considered a two lead or equivalently a two well system. Therefore if the number of leads are $N$ then ${\rho}_{sc}$  will be an $N \times N $ matrix. In brief ${\rho}_{sc}$ is like a ``{\bf Matrix representation of Number operator}, which is non-diagonal in spatial(lead) as well in spin space.

The formulation of the theory in terms of lead indices(or in mesoscopic language) is helpful to connect it with the experimentally measured local spin-polarization in the leads as well in the sample. In-fact the Magneto-optical methods such as Faraday rotation and Kerr spectroscopy allows one to probe the local spin polarization along various points in the leads as well in the sample\cite{Meier,rizo,Kikkawa,Fabian},which can be directly related to the spin-density polarization vector calculated in this article. 

In summary we have present a formulation of {\bf Quantum Non-Abelian} hydrodynamics. We have show that it is physically equivalent to having 
spin-space entanglement non-zero, Non-unitarity of scattering matrix and the quasi particle excitations are Anyonic. Moreover we have shown that in presence of spin-dependent potentials, the quantum non-abelian hydrodynamics is a physical solution which arise naturally, while abelian hydrodynamics which is equivalent to having unitary scattering matrix turns out to be a mathematically forced solution.  

Possible future direction would be of two kinds, namely, extending the  hydrodynamic approach for the noise spectra of currents which is expected to show non-trival statistical correlation and are easily accessible through experiments. The second approach would be to relax the hydrodynamic approximation and build a local non-adiabatic theory, which would be rather involved mathematically. However we feel that a path integral approach 
of quantum mechanics would be most suitable and economical mathematical toll for this purpose\cite{Feynman}

\appendix
\section{Appendix}
\label{appendix1}
%\arabic{0}
\renewcommand{\theequation}{S\arabic{equation}}
\setcounter{equation}{0}

%A scalar function $f(\bm{r})\equiv f(x,y,z)$ of Cartesian coordinates $r=(x,y,z)$, should be understood as a function of three scalar variables $\{x,y,z\}$. While $r=\sqrt(x^{2}+y^{2}+z^{2})$ is distance from the origin of the coordinate systems.

\subsection{Identities involving Vectorial Matrices}
\label{sec-S1}
In the expression (\ref{eq:NC-2}) of main text if we take $\bm{B}=\bm{A}^{\dagger}$, then the dagger operation of cross product is,
\begin{eqnarray}
(\bm{A}\times\bm{A}^{\dagger})^{\dagger}&=& \sum_{ijk}\varepsilon_{ijk}(A_{j}A_{k}^{\dagger})^{\dagger}\hat{i} \nonumber \\
&=&  \sum_{ijk}\varepsilon_{ijk}(A_{k}A_{j}^{\dagger})\hat{i}\equiv  -\sum_{ijk}\varepsilon_{ikj}(A_{k}A_{j}^{\dagger})\hat{i} \nonumber \\
&=& - (\bm{A}\times\bm{A}^{\dagger}),
\label{eq:S1}
\end{eqnarray}
therefore, $\{i(\bm{A}\times\bm{A}^{\dagger})\}^{\dagger}= \{i(\bm{A}\times\bm{A}^{\dagger})\}$ is hermitian. This is an important identity which we have used frequently for vectorial operators. Similarly the scalar triple product of vectorial matrices $\bm{A}$,$\bm{B}$ and $\bm{C}$ is,
\begin{equation}
 (\bm{A}\times \bm{B})\cdot \bm{C}=\sum_{ijk}\varepsilon_{ijk}A_{j}B_{k}C_{i}=\sum_{ijk}\varepsilon_{ijk}A_{j}B_{k}C_{i} \label{eq:S2}
\end{equation}

The scalar triple product of three vectorial matrices $\bm{A}$,$\bm{A}^{\dagger}$ and $\bm{B}$,is equal to the vectorial matrix, $(\bm{A}\times\bm{B})\cdot\bm{A}^{\dagger}$, we are interested in the Hermitian conjugate of this matrix which is,
\begin{eqnarray}
 \{(\bm{A}\times\bm{B})\cdot\bm{A}^{\dagger}\}^{\dagger}&=&\sum_{ijk}\varepsilon_{ijk}(A_{j}B_{k}A^{\dagger}_{i})^{\dagger} \nonumber \\
&=&\sum_{ijk}\varepsilon_{ijk}(A_{i}B_{k}^{\dagger}A_{j}^{\dagger}) \nonumber \\
&=& -\sum_{ijk}\varepsilon_{ikj}(A_{i}B_{k}^{\dagger}A_{j}^{\dagger}) \nonumber \\
&=&-(\bm{A}\times\bm{B}^{\dagger})\cdot\bm{A}^{\dagger}. \label{eq:S3}
\end{eqnarray}
From the above expression it is clear that if $\bm{B}$ is Hermitian,i.e.,  $\bm{B}^{\dagger}=\bm{B}$, in that case the scalar triple
product defined above in anti-hermitian. Hence if we take $\bm{A}=\bm{S}$ and $\bm{B}=\bm{\mathcal{\varrho}}$, we see that
the term $i(\bm{S}\times\bm{\varrho})\cdot \bm{S}^{\dagger}$, which is the last term in the first line of scattered 
density matrix is also hermitian.

Using defining eq.(\ref{eq:NC-5}) from the main text we arrive at the following three most important identities involving {\bf vectorial matrices} and {\bf Pauli vector},
\begin{eqnarray}
 (\bm{\sigma}\cdot \bm{A})(\bm{\sigma}\cdot \bm{B})= \sum_{ij}(\sigma_{i}A_{i})(\sigma_{j}B_{j})&=&\sum_{ij}(\sigma_{i}\sigma_{j})(A_{i}B_{j}) \nonumber \\
&=&\sum_{ij}(\delta_{ij}\bm{I}_{2}^{s}+i\epsilon_{ijk}\sigma_{k})(A_{i}B_{j}) \nonumber \\
&=&\sum_{i}(A_{i}B_{i})\bm{I}_{2}^{s}+i\sum_{ij}\epsilon_{ijk}(A_{i}B_{j})\sigma_{k} \label{eq:S4} \\
(\bm{\sigma}\cdot \bm{A})(\bm{\sigma}\cdot \bm{B})&=&(\bm{A}\cdot\bm{B})\bm{I}^{s}+i(\bm{A}\times\bm{B})\cdot \bm{\sigma} \label{eq:S5},
\end{eqnarray}
Note that in eq.(\ref{eq:S4}) above $A_{i}$ and $B_{i}$ are matrices and since matrix product is non-commutative hence the ordering of vectorial matrices in the compact expression (\ref{eq:S5}) has to be same. Taking care of this ordering we obtain the remaining three identities,
\begin{eqnarray}
%( \bm{\sigma}\cdot \bm{A}) ( \bm{\sigma}\cdot \bm{B})=(\bm{A}\cdot\bm{B})+i\bm{\sigma}\cdot (\bm{A}\times \bm{B}). \label{eq:S7} \\
 ( \bm{\sigma}\cdot \bm{B}) ( \bm{\sigma}\cdot \bm{A})=(\bm{B}\cdot\bm{A})+i\bm{\sigma}\cdot (\bm{B}\times \bm{A}) \neq 
( \bm{\sigma}\cdot \bm{A}) ( \bm{\sigma}\cdot \bm{B}), \label{eq:S6} \\
( \bm{\sigma}\cdot \bm{A}) ( \bm{\sigma}\cdot \bm{A})=(\bm{A}\cdot\bm{A})+i\bm{\sigma}\cdot (\bm{A}\times \bm{A}) \label{eq:S7}
\end{eqnarray}
Notice that when $\bm{A}$ and $\bm{B}$ are simple vectors, the dot product term in
eq.(\ref{eq:S1}) is commutative while for vectorial matrices this is in general not true because matrix product is non-commutative, therefore, the identities given above in eqs.(\ref{eq:S5}-\ref{eq:S7}) are qualitatively different from the standard text book identities\cite{}
involving simple vectors. Moreover for vectorial matrices both terms in eq.(\ref{eq:S5}) are non-commutative hence care should be taken while manipulating such terms. The three identities eqs.(\ref{eq:S5}-\ref{eq:S7}) can be combined into
a single expression as,
\begin{equation}
 [(\bm{\sigma}\cdot \bm{A}) ( \bm{\sigma}\cdot \bm{B}),( \bm{\sigma}\cdot \bm{B}) ( \bm{\sigma}\cdot \bm{A})]_{-}=[ (\bm{A}\cdot \bm{B}),(\bm{B}\cdot \bm{A})]_{-}+i\bm{\sigma}[(\bm{A}\times \bm{B}),(\bm{B}\times \bm{A})]_{-} \label{eq:S8}
\end{equation}
where $[..., ...]_{-}$ stands for anti-commutator.

\subsection{Scalar and Vectorial Traces of density matrix}
\label{S1-1}
We note that the incident density matrix(set of eqs.(\ref{eq:IDM-2}-\ref{eq:IDM-3}) is composed of a scalar and a vectorial part and the same will be true for scattered density matrix which appears in the set of eqs.(\ref{eq:SDM-2}-\ref{eq:SDM-3}) and both have a general structure 
\begin{equation}
\rho_{a}= \varrho_{a,0}\otimes \bm{I}_{2}^{s}+\bm{\varrho}_{a}\cdot \bm{\sigma}\equiv \varrho_{sc,0}\otimes \bm{I}_{2}+\sum_{i}\varrho_{a,i}\otimes \sigma_{i} \,\,\, a \in\{in, sc\}
\end{equation}
where  subscript $a \in\{in, sc\}$ corresponding to incident or scattered density matrix and  $\varrho_{a,0}$ and $\bm{\varrho}_{a}$ are scalar and vectorial matrices in the lead space while  $\bm{I}_{2}$ and  $\bm{\sigma}$ are identity matrix and Pauli vectorial matrix in the spin space respectively. The number density and spin density for the above density matrix is given by the scalar and vectorial trace operations
as follows,
\begin{eqnarray}
 \text{Tr}[\rho_{a}]=\text{Tr}[(\varrho_{a,0}\otimes \bm{I}_{2}^{s}+\sum_{i}\varrho_{a,i}\otimes \sigma_{i})] =2\text{tr}[\varrho_{a,0}]=\sum_{\alpha}\varrho_{a,0,\alpha\alpha}
\label{eq:S1-1-1}
\end{eqnarray}
\begin{eqnarray}
 \text{Tr}[\rho_{a}\bm{\sigma}]=\text{Tr}[(\varrho_{a,0}\otimes \bm{I}_{2}^{s}+\sum_{i}\varrho_{a,i}\otimes \sigma_{i})(\sum_{j}\sigma_{j}\hat{j})] 
=\text{Tr}[\sum_{ij}\varrho_{a,i}\otimes \sigma_{i}\sigma_{j}\hat{j}] \nonumber \\
=\text{Tr}[\sum_{ij}\varrho_{a,i}\otimes\left\{ i\varepsilon_{ijk}\sigma_{k}+\delta_{ij}\otimes \bm{I}_{2}^{s}\right\}\hat{j}] 
=\text{Tr}[\sum_{i}\varrho_{a,i}\hat{i}\otimes \bm{I}_{2}^{s}] = \text{Tr}[\bm{\varrho}_{a}\cdot \bm{I}_{2}^{s}]\nonumber\\
=2\text{tr}[\bm{\varrho}_{a}]=\sum_{i\alpha}(\varrho_{a,i})_{\alpha\alpha}\hat{i} \label{eq:S1-1-2}
\end{eqnarray}

\subsection{Calculation of $\varrho_{sc}$}
\label{sec-S2}
In this section we provide the details of the intermediate steps which leads to the simplified expression (\ref{eq:SDM-2}) from expression (\ref{eq:SDM-1}).
Toward this end we see that the following expression gets simplified to,
\begin{eqnarray}
 [{S}_{0}{\bm{I}}_{2}+\bm{S}\cdot\bm{\sigma}][\rho_{in}\bm{I}_{2}+{\bm{\rho}}_{in}\cdot \bm{\sigma}]&=&[{S}_{0}\rho_{0}+\bm{S}\cdot{\bm{\rho}}_{in}]\bm{I}_{2} \nonumber \\
&+&[({S}_{0}{\bm{\rho}}_{in}+\bm{S}\rho_{in})+\bm{\dot{\iota}}(\bm{S}\times{\bm{\rho}}_{in})]\cdot \bm{\sigma} \label{eq:S9},
\end{eqnarray}
where we have made use of the identities (\ref{eq:S5}-\ref{eq:S7}).
The above expression multiplied from right by $[{S}_{0}^{\dagger}{\bm{I}}_{2}+\bm{S}^{\dagger}\cdot\bm{\sigma}]$ is the scattered density matrix given by  (\ref{eq:SDM-1}). To simplify this multiplication we look at various terms separately,
\begin{eqnarray}
[({S}_{0}\rho_{in}+\bm{S}\cdot{\bm{\rho}}_{in})\bm{I}_{2}] [{S}_{0}^{\dagger}{\bm{I}}_{2}+\bm{S}^{\dagger}\cdot\bm{\sigma}]&=& [{S}_{0}\rho_{in}{S}_{0}^{\dagger} 
+(\bm{S}\cdot{\bm{\rho}}_{in}){S}_{0}^{\dagger}]\bm{I}_{2} \nonumber \\ 
&+& [{S}_{0}\rho_{in}\bm{S}^{\dagger}+(\bm{S}\cdot{\bm{\rho}}_{in})\bm{S}^{\dagger}] \cdot \bm{\sigma}.\label{eq:S10}
\end{eqnarray}
For the remaining terms we proceed as follows,
\begin{eqnarray}
[( {S}_{0}{\bm{\varrho}}_{in}+\bm{S}\rho_{in})\cdot \bm{\sigma}] [{S}_{0}^{\dagger}{\bm{I}}_{2}+\bm{S}^{\dagger}\cdot\bm{\sigma}]&=&({S}_{0}{\bm{\rho}}_{in}\cdot\bm{S}^{\dagger}+\bm{S}\rho_{in}\cdot\bm{S}^{\dagger}){\bm{I}}_{2} \nonumber \\
&+& [ ({S}_{0}{\bm{\rho}}_{in}{S}_{0}^{\dagger}+\bm{S}\rho_{in}{S}_{0}^{\dagger})]\cdot \bm{\sigma} \nonumber \\
&+& \bm{\dot{\iota}} [{S}_{0}({\bm{\rho}}_{in}\times\bm{S}^{\dagger})+(\bm{S}\rho_{in}\times\bm{S}^{\dagger})]\cdot \bm{\sigma}
\label{eq:S11}
\end{eqnarray} 
\begin{eqnarray}
[\bm{\dot{\iota}}(\bm{S}\times{\bm{\rho}}_{in})][{S}_{0}^{\dagger}{\bm{I}}_{2}+\bm{S}^{\dagger}\cdot\bm{\sigma}]&=&\bm{\dot{\iota}}[(\bm{S}\times{\bm{\rho}}_{in})\cdot\bm{S}^{\dagger}]{\bm{I}}_{2} +\bm{\dot{\iota}}[(\bm{S}\times{\bm{\rho}}_{in}){S}_{0}^{\dagger}]\cdot\bm{\sigma} \nonumber \\
&-&[(\bm{S}\times{\bm{\rho}}_{in})\times \bm{S}^{\dagger} ]\cdot\bm{\sigma} \label{eq:S12}.
\end{eqnarray}
Adding expressions (\ref{eq:S10}),(\ref{eq:S11}) and (\ref{eq:S12}) we obtained the scattered density matrix given by eq.(\ref{eq:SDM-2}).

\subsection{Hermiticity of Scattered density matrix}
\label{sec-S3}
We discusses the hermiticity of different terms of expression (\ref{eq:SDM-2})  given in the main text. Toward this end we note that for incident density matrix given by eq.(\ref{eq:IDM-3}) in main text, $\rho_{in}$ and $\bm{\varrho}$ are scalar and vectorial diagonal matrices (see eq.(\ref{eq:IDM-2}-\ref{eq:IDM-3}) in main text), therefore, $\rho_{in}^{\dagger}=\rho_{in}$ and $\bm{\varrho}^{\dagger}=\bm{\varrho}$. 

The scalar part of the scattered density matrix is, (see eqs.(\ref{eq:SDM-2}) and (\ref{eq:SDM-3}) in main text)
\begin{equation}
{\varrho}_{sc,0}= [S_{0}\rho_{in}S_{0}^{\dagger}+\{(\bm{S}\cdot{\bm{\rho}}_{in})S_{0}^{\dagger}+
S_{0}({\bm{\rho}}_{in}\cdot\bm{S}^{\dagger})\}+
(\bm{S}\rho_{in}\cdot\bm{S}^{\dagger})+i(\bm{S}\times{\bm{\rho}}_{in})\cdot \bm{S}^{\dagger}]\otimes \bm{I}_{2}  \nonumber 
\end{equation}
In the above expression first term, $[S_{0}\rho_{in}S_{0}^{\dagger}]^{\dagger}=[S_{0}\rho_{in}S_{0}^{\dagger}]$ is
obviously hermitian. The second and third term are together hermitian as is shown below,
\begin{eqnarray}
\{(\bm{S}\cdot{\bm{\rho}}_{in})S_{0}^{\dagger}+
S_{0}({\bm{\rho}}_{in}\cdot\bm{S}^{\dagger})\}^{\dagger}&=&\left\{\sum_{i}(S_{i}\varrho_{i}S_{0}^{\dagger})^{\dagger}+\sum_{i}(S_{0}\varrho_{i}S_{i}^{\dagger})^{\dagger}\right\} \nonumber \\
&=&\{S_{0}({\bm{\rho}}_{in}\cdot\bm{S}^{\dagger})+(\bm{S}\cdot{\bm{\rho}}_{in})S_{0}^{\dagger}\}. \label{eq:S13}
\end{eqnarray}
The hermiticity of the fourth term also follows similarly, i.e., 
$(\bm{S}\rho_{in}\cdot\bm{S}^{\dagger})^{\dagger}=\sum_{i}(S_{i}\rho_{in}S_{i}^{\dagger})^{\dagger}=
(\bm{S}\rho_{in}\cdot\bm{S}^{\dagger})$. 
To show that fifth term is also hermitian requires some manipulation such as,
\begin{eqnarray}
 [\bm{\dot{\iota}}(\bm{S}\times{\bm{\rho}}_{in})\cdot \bm{S}^{\dagger}]^{\dagger}&=&-\bm{\dot{\iota}}[\sum_{i}(\bm{S}\times{\bm{\rho}}_{in})_{i} {S}^{\dagger}_{i}]^{\dagger} \equiv -\bm{\dot{\iota}}[\sum_{ijk}\varepsilon_{ijk}S_{j}{\rho}_{in,k}{S}^{\dagger}_{i}]^{\dagger}  \nonumber \\
&=&\bm{\dot{\iota}}[\sum_{ijk}\varepsilon_{ikj}S_{i}{\rho}_{in,k}{S}^{\dagger}_{j}]\equiv [\bm{\dot{\iota}}(\bm{S}\times{\bm{\rho}}_{in})\cdot \bm{S}^{\dagger}], \label{eq:S14}
\end{eqnarray}
where we have made use of the fact that $\varepsilon_{ijk}=-\varepsilon_{ikj}$ and $\rho_{in,k}^{\dagger}=\rho_{in,k}$ in the second line of above
expression. Therefore we have shown the hermiticity of different terms of scalar part of scattered density matrix, i.e.,$\rho_{sc}$ or equivalently of the coefficient of $\bm{I}_{2}^{s}$ in eqs.(\ref{eq:SDM-2}) in main text. Moreover we also note that the expression (\ref{eq:S14}) can equivalently be expressed as,
Toward this end we first note that $[(\bm{S}\times{\bm{\rho}}_{in})\cdot \bm{S}^{\dagger}]$ can be written in a compact form as,
\begin{eqnarray}
[\bm{\dot{\iota}}(\bm{S}\times{\bm{\rho}}_{in})\cdot \bm{S}^{\dagger}]&=&\bm{\dot{\iota}}[(S_{y}\rho_{in,z}-S_{z}\rho_{y})S_{x}^{\dagger}+(S_{z}\rho_{in,x}-S_{x}\rho_{in,z})S_{y}^{\dagger} +(S_{x}\rho_{y}-S_{y}\rho_{in,x})S_{z}^{\dagger}] \nonumber \\
&=&\bm{\dot{\iota}}[(S_{z}\rho_{in,x}S_{y}^{\dagger}-S_{y}\rho_{in,x}S_{z}^{\dagger})+ (S_{x}\rho_{in,y}S_{z}^{\dagger}-S_{z}\rho_{in,y}S_{x}^{\dagger})+(S_{y}\rho_{in,z}S_{x}^{\dagger}-S_{x}\rho_{in,z}S_{y}^{\dagger})] \nonumber \\
&=&\sum_{(ijk)_{c}}\bm{\dot{\iota}}(S_{k}\rho_{in,i}S_{j}^{\dagger}-S_{j}\rho_{in,i}S_{k}^{\dagger}) \label{eq:S15}
\end{eqnarray}
where the summation symbol the notation $(ijk)_{c}$ implies that only cyclic permutations of ${x, y, z}$ is allowed, i.e, $\sum_{(ijk)_{c}}\equiv [(xyz), (yzx), (zxy)]$.

Now let us look at the vectorial part of scattered density matrix ${\bm{\rho}}_{sc}$ in eq.(\ref{eq:SDM-3}) or equivalently coefficient of $\bm{\sigma}$ in the expression  (\ref{eq:SDM-2}) which is,
\begin{eqnarray}
 \bm{\varrho}_{sc}&=&[S_{0}{\bm{\rho}}_{in}S_{0}^{\dagger}+\{S_{0}\rho_{in}\bm{S}^{\dagger}
+\bm{S}\rho_{in}S_{0}^{\dagger}\}+\{(\bm{S}\cdot{\bm{\rho}}_{in})\bm{S}^{\dagger}
-(\bm{S}\times {\bm{\rho}}_{in})\times \bm{S}^{\dagger})\}]\cdot \bm{\sigma} \nonumber \\
&+&[\{\bm{\dot{\iota}}S_{0}({\bm{\rho}}_{in}\times \bm{S}^{\dagger})+\bm{\dot{\iota}}(\bm{S}\times{\bm{\rho}}_{in})S_{0}^{\dagger}\} 
+\bm{\dot{\iota}}(\bm{S}\rho_{in}\times \bm{S}^{\dagger}) ]\cdot\bm{\sigma}.\label{eq:S16}
\end{eqnarray}
In the above expression the first terms is obviously hermitian while the second and third terms are together hermitian as can be seen easily by
following similar step as was done previously. To show that the fourth and fifth terms are together hermitian we proceed as follows. First we note that the fourth and fifth term can be written in a simplified form as,
\begin{eqnarray}
(\bm{S}\cdot{\bm{\rho}}_{in})\bm{S}^{\dagger}=\sum_{i}(\sum_{j}S_{j}\rho_{in,j})S^{\dagger}_{i}
\hat{i}
=\sum_{i,j\neq i}S_{j}\rho_{in,j}S^{\dagger}_{i}\hat{i}
+\sum_{i}S_{i}\rho_{in,i}S^{\dagger}_{i}\hat{i} \label{eq:S17}
\end{eqnarray}
\begin{eqnarray}
\{(\bm{S}\times {\bm{\rho}}_{in})\times \bm{S}^{\dagger}\}=\sum_{ijk}\varepsilon_{ijk}
(\bm{S}\times {\bm{\rho}}_{in})_{j}S^{\dagger}_{k}\hat{i} =\sum_{ijk,lm}\varepsilon_{ijk}\varepsilon_{jlm}S_{l}\rho_{in,m}S^{\dagger}_{k}\hat{i}\nonumber \\
=-\sum_{ik,lm}\varepsilon_{jik}\varepsilon_{jlm}S_{l}\rho_{in,m}S^{\dagger}_{k}\hat{i}=-\sum_{i k,lm}(\delta_{il}\delta_{km}-\delta_{im}\delta_{kl})S_{l}\rho_{in,m}S^{\dagger}_{k}\hat{i} \nonumber \\
=\sum_{i,k\neq i}S_{k}\rho_{in,i}S^{\dagger}_{k}\hat{i}-S_{i}\rho_{in,k}S^{\dagger}_{k}\hat{i} 
\equiv \sum_{i,j\neq i}S_{j}\rho_{in,i}S^{\dagger}_{j}\hat{i}-S_{i}\rho_{in,j}S^{\dagger}_{j}\hat{i} \label{eq:S18}
%&=&\sum_{j,k\neq j}\left[S_{j}\varrho_{k}S_{j}^{\dagger}\hat{k}-
%S_{j}\varrho_{k}S_{k}^{\dagger}\hat{j}\right]\label{eq:S-h6}.
\end{eqnarray}
Using eqs.(\ref{eq:S17},\ref{eq:S18}) we can see that the following combined expressions is explicitly hermitian, i.e., 
\begin{eqnarray}
 \{(\bm{S}\cdot{\bm{\rho}}_{in})\bm{S}^{\dagger}
-(\bm{S}\times {\bm{\rho}}_{in})\times \bm{S}^{\dagger})\}=\sum_{i,j,j\neq j}\{(S_{j}\rho_{in,j}S^{\dagger}_{i}+S_{i}\rho_{in,j}S_{j}^{\dagger})\hat{i}\} \nonumber \\
+\sum_{i}S_{i}\rho_{in,i}S^{\dagger}_{i}\hat{i} 
-\sum_{i,j ,j\neq i}S_{j}\rho_{in,i}S_{j}^{\dagger}\hat{i} \nonumber \\
\equiv \sum_{(ijk)_{c}}(S_{i}\rho_{in,j}S_{j}^{\dagger}+S_{i}\rho_{in,k}S_{k}^{\dagger}+S_{j}\rho_{in,j}S_{i}^{\dagger}+S_{k}\rho_{in,k}S_{i}^{\dagger})\hat{i} \nonumber \\
+ \sum_{(ijk)_{c}}(S_{i}\rho_{in,i}S_{i}^{\dagger}-S_{j}\rho_{in,i}S_{j}^{\dagger}-S_{k}\rho_{in,i}S_{k}^{\dagger})\hat{i}
\label{eq:S19}
\end{eqnarray}
The remaining terms in second line of the expression (\ref{eq:S16}) can be written explicitly as,
\begin{eqnarray}
 \{\bm{\dot{\iota}}S_{0}({\bm{\rho}}_{in}\times \bm{S}^{\dagger})+
\bm{\dot{\iota}}(\bm{S}\times{\bm{\rho}}_{in})S_{0}^{\dagger}\} &=&\sum_{(ijk)_{c}}[\bm{\dot{\iota}}(S_{0}\rho_{in,j}S_{k}^{\dagger}-S_{k}\rho_{in,j}S_{0}^{\dagger})\nonumber \\
&+&\bm{\dot{\iota}}(S_{j}\rho_{in,k}S_{0}^{\dagger}-S_{0}\rho_{in,k}S_{j}^{\dagger})]\hat{i} \label{eq:S20} \\
\bm{\dot{\iota}}(\bm{S}\rho_{in}\times \bm{S}^{\dagger})&=&\{\sum_{\{ijk\}}\bm{\dot{\iota}}(S_{j}\rho_{in}S_{k}^{\dagger}-S_{k}\rho_{in}S_{j}^{\dagger})\hat{i}\},\label{eq:S21}
\end{eqnarray}
as is seen the above two expressions are manifestly hermitian. 
Therefore the vectorial part of scattered density matrix is also hermitian.

\subsection{Trace calculation of scattered density matrix: ``Number density and Spin Polarization of scattered fluid''}
\label{sec-S4}
In the section of supplementary information we provide details of the steps which leads to the simplified expressions (\ref{eq:DP-1}) and (\ref{eq:DP-SD}) in section(\ref{sec-Num-DP}) of the main text. Towards this end we need to calculate $\text{tr}[\rho_{sc}]$ and $\text{tr}[\bm{\rho}_{sc}]$. Evaluating these two traces requires calculating traces for the following expressions,
\begin{eqnarray}
\text{tr}[\rho_{sc,0}]&=& \sum_{\alpha}\rho_{sc,\alpha\alpha,0} \nonumber \\
&=&\sum_{\alpha}[S_{0}\rho_{in}S_{0}^{\dagger}+(\bm{S}\rho_{in}\cdot\bm{S}^{\dagger})]_{\alpha\alpha} \nonumber \\
&+&
\sum_{\alpha\alpha}[
\{(\bm{S}\cdot{\bm{\rho}}_{in})S_{0}^{\dagger}+
S_{0}({\bm{\rho}}_{in}\cdot\bm{S}^{\dagger})\}+i(\bm{S}\times{\bm{\rho}}_{in})\cdot \bm{S}^{\dagger}]_{\alpha\alpha}\label{eq:S22} 
\end{eqnarray}
\begin{eqnarray}
\text{tr}[\bm{\varrho}_{sc}]&=&\sum_{i}\varrho_{sc,i,\alpha\alpha}\hat{i} \nonumber \\
&=&\sum_{\alpha}(S_{0}{\bm{\rho}}_{in}S_{0}^{\dagger}+\{S_{0}\rho_{in}\bm{S}^{\dagger}
+\bm{S}\rho_{in}S_{0}^{\dagger}\}+\{(\bm{S}\cdot{\bm{\rho}}_{in})\bm{S}^{\dagger}
-(\bm{S}\times {\bm{\rho}}_{in})\times \bm{S}^{\dagger})\})_{\alpha\alpha}\nonumber \\
&+&\sum_{\alpha}(\{ \bm{\dot{\iota}}S_{0}({\bm{\rho}}_{in}\times \bm{S}^{\dagger})+\bm{\dot{\iota}}(\bm{S}\times{\bm{\rho}}_{in})S_{0}^{\dagger}\} 
+\bm{\dot{\iota}}(\bm{S}\rho_{in}\times \bm{S}^{\dagger}))_{\alpha\alpha}\label{eq:S23}.
\end{eqnarray} 
where, $\text{tr}$, implies trace over Cartesian,i.e, (x,y,z), and the lead index, i.e., $\alpha,\beta$.
Before we proceed further it is helpful to note that the {\it scalar}($\rho_{in}$) and {\bf vectorial} part($\bm{\rho_{in}}=\sum_{i}\rho_{in,i}\hat{i}$) of incident density matrix is diagonal in lead basis, therefore matrix elements of matrices $\rho_{in}$ and $\rho_{in,i}$ are given by,   
\begin{eqnarray}
\rho_{in,\gamma\beta}=\rho_{in,\beta,0}\delta_{\gamma\beta} \label{eq:S23-1}\\
{\rho}_{in,\gamma\beta,i}={p}_{in,\beta,i}\delta_{\gamma\beta} \label{eq:S23-2}
\end{eqnarray}
where we have made use of the defining equations (\ref{eq:IDM-1}-\ref{eq:IDM-3}) in the section(\ref{sub-sec-R-IDM}) of main text. We will be using these two relation frequently to simplify the trace expression.

Now let us look at the first terms of expression (\ref{eq:S22}),
\begin{eqnarray}
\sum_{\alpha}(S_{0}\rho_{in}S_{0}^{\dagger})_{\alpha\alpha}&=&\sum_{\alpha,\beta}(S_{0}\rho_{in})_{\alpha\beta}(S_{0}^{\dagger})_{\beta\alpha}=\sum_{\alpha,\beta,\gamma}S_{\alpha\gamma,0}\rho_{in,\gamma\beta}S_{\alpha\beta,0}^{*}  \nonumber \\
&=&\sum_{\alpha,\beta,\gamma}S_{\alpha\gamma,0}\rho_{in,\beta,0}\delta_{\gamma\beta}S_{\alpha\beta,0}^{*}\equiv\sum_{\alpha,\beta}|S_{\alpha\beta,0}|^{2}\rho_{in,\beta,0}.\label{eq:S24}
\end{eqnarray}
Similarly the second term in eq.(\ref{eq:S22}) above simplifies to,
\begin{eqnarray}
\sum_{\alpha}(\bm{S}\rho_{in}\cdot\bm{S}^{\dagger})_{\alpha\alpha}=\sum_{i,\alpha}(S_{i}\rho_{in}{S}^{\dagger}_{i})_{\alpha\alpha}=\sum_{i,\alpha\gamma}(S_{i}\rho_{in})_{\alpha\gamma}({S}^{\dagger}_{i})_{\gamma\alpha} \nonumber \\
=\sum_{i,\alpha\beta\gamma}S_{\alpha\beta,i}\rho_{in,\gamma\beta}{S}^{*}_{\alpha\gamma,i}=\sum_{i,\alpha\beta\gamma}S_{\alpha\beta,i}{S}^{*}_{\alpha\gamma,i}\rho_{in,\beta,0}\delta_{\gamma \beta}=
\sum_{i,\alpha,\beta}|S_{\alpha\beta,i}|^{2}\rho_{in,\beta,0}. \label{eq:S25}
\end{eqnarray}
Adding expressions (\ref{eq:S24}) and (\ref{eq:S25}), we obtain 
\begin{equation}
\sum_{\alpha} [(S_{0}\rho_{in}S_{0}^{\dagger}) +(\bm{S}\rho_{in}\cdot\bm{S}^{\dagger})]_{\alpha\alpha}=\sum_{\alpha,\beta}[|S_{\alpha\beta,0}|^{2}+\sum_{i}|S_{\alpha\beta,i}|^{2}]\rho_{in,\beta,0}\equiv \sum_{\alpha,\beta} [\mathcal{C}_{0,\alpha\beta}]\rho_{in,\beta,0}\label{eq:S26}
\end{equation}
which is same as first term of eq.(\ref{eq:DP-1}) in main text.
The trace of third and fourth terms of expression (\ref{eq:S22}) is,
\begin{eqnarray}
\sum_{\alpha}[ \{(\bm{S}\cdot{\bm{\rho}}_{in})S_{0}^{\dagger}+
S_{0}({\bm{\rho}}_{in}\cdot\bm{S}^{\dagger})\}]_{\alpha\alpha}&=&\sum_{i,\alpha}[S_{i}\rho_{in,i}S_{0}^{\dagger}+S_{0}\rho_{in,i}S_{i}^{\dagger}]_{\alpha\alpha} \nonumber \\
&=&\sum_{i}\sum_{\alpha\beta}[S_{\alpha\beta,i}\rho_{in,\beta,i}S_{\alpha\beta,0}^{*}+S_{\alpha\beta,0}\rho_{in,\beta,i}S_{\alpha\beta,i}^{*}]\nonumber \\
&\equiv& \sum_{i,\alpha\beta}2[\text{Re}(S_{\alpha\beta,0}S_{\alpha\beta,i}^{*})] p_{in,\beta,i}.\label{eq:S27}
\end{eqnarray}
Similarly the trace of last terms of expression (\ref{eq:S22}) can be calculated using the explicit form given in eq.(\ref{eq:S15}) which leads to,
\begin{eqnarray}
\sum_{\alpha}[ \bm{\dot{\iota}}(\bm{S}\times{\bm{\rho}}_{in})\cdot \bm{S}^{\dagger}]_{\alpha\alpha}&=&
\sum_{(ijk)_{c},\alpha}\bm{\dot{\iota}}[(S_{k}\rho_{in,i}S^{\dagger}_{j}-S_{j}\rho_{in,i}S^{\dagger}_{k})]_{\alpha\alpha}\nonumber \\
&\equiv&-\sum_{(ijk)_{c},\alpha\beta}2\text{Im}[S_{\alpha\beta,j}^{*}S_{\alpha\beta,k}]p_{in,\beta,i} \label{eq:S28}.
\end{eqnarray}
where we have made use of the relations eqs.(\ref{eq:S23-1}-\ref{eq:S23-2}).
Adding expression (\ref{eq:S26}), (\ref{eq:S27}) and (\ref{eq:S28}) we obtain,
\begin{eqnarray}
\text{tr}[\rho_{sc,0}]&=& \sum_{\alpha,\beta} [|S_{\alpha\beta,0}|^{2}+\sum_{i}|S_{\alpha\beta,i}|^{2}]\rho_{in,\beta,0} \nonumber \\
&+&\sum_{(ijk)_{c}}
2[ \text{Re}(S_{\alpha\beta,i}S_{\alpha\beta,i}^{*})-\text{Im}(S_{\alpha\beta,j}^{*}S_{\alpha\beta,k})] p_{in,\beta,i} \nonumber \\
&\equiv&  \sum_{\alpha,\beta} [\mathcal{C}_{0,\alpha\beta}\rho_{in,\beta,0}+2 \sum_{(ijk)_{c}}{\cal{Q}}^{\alpha\beta}_{0ijk}p_{in,\beta,i}]
\label{eq:S28-1}
\end{eqnarray}
which is same as the eq.(\ref{eq:DP-1}) of the main text.

Now we evaluate,$\text{tr}[{\bm{\varrho}}_{sc}]$,i.e, the expression (\ref{eq:S23}). Toward this end  we note that first three terms in this expression simplifies to,
\begin{eqnarray}
 \sum_{\alpha}[S_{0}{\bm{\rho}}_{in}S_{0}^{\dagger}]_{\alpha\alpha}=\sum_{i,\alpha\beta}|S_{\alpha\beta,0}|^{2}p_{in,\beta,i} \hat{i}\label{eq:S29}\\
\text{tr}[\{S_{0}\rho_{in}\bm{S}^{\dagger}
+\bm{S}\rho_{in}S_{0}^{\dagger}\}]=\sum_{i,\alpha\beta}2\text{Re}(S_{\alpha\beta,0}S_{\alpha\beta,i}^{*})\rho_{in,\beta,0}\hat{i} \label{eq:S30}
\end{eqnarray}
while the trace of fourth and fifth terms can be evaluated using the explicit expression(\ref{eq:S19}) which leads to,
\begin{eqnarray}
\text{tr}\{(\bm{S}\cdot{\bm{\rho}}_{in})\bm{S}^{\dagger}
-(\bm{S}\times {\bm{\rho}}_{in})\times \bm{S}^{\dagger})\}&=&\sum_{i,j,j\neq i,\alpha\beta}\left\{(S_{\alpha\beta,j}\rho_{in,\beta,j}S^{*}_{\alpha\beta,i}+S_{\alpha\beta,i}\rho_{in,\beta,j}S_{\alpha\beta,j}^{*}) \nonumber \right. \\
&+&\left. (S_{\alpha\beta,i}\varrho_{i,\beta\beta}S^{*}_{\alpha\beta,i}
-S_{\alpha\beta,j}\rho_{in,\beta,i}S_{\alpha\beta,j}^{*})\right\}\hat{i} \nonumber \\
 &=&\sum_{i,j,j\neq i,\alpha\beta} \left\{ 2\text{Re}(S_{\alpha\beta,i}S_{\alpha\beta,j}^{*})\rho_{in,\beta,j} \nonumber \right. \\
&+&\left.|S_{\alpha\beta,i}|^{2}\varrho_{i,\beta\beta} 
- |S_{\alpha\beta,j}|^{2}\rho_{in,\beta,i}\right\}\hat{i} \nonumber \\
&\equiv& \sum_{(ijk)_{c},\alpha\beta} \left.\Bigg\{ (|S_{\alpha\beta,i}|^{2}
- |S_{\alpha\beta,j}|^{2}-|S_{\alpha\beta,k}|^{2})p_{in,\beta,i} \nonumber \right. \\
&+&\left. 2\text{Re}(S_{\alpha\beta,i}S_{\alpha\beta,j}^{*})p_{in,\beta,j} \nonumber \right. \\
&+&\left. 2\text{Re}(S_{\alpha\beta,i}S_{\alpha\beta,k}^{*})p_{in,\beta,k} \right.\Bigg\} \hat{i}\label{eq:S31} ,
\end{eqnarray}
Using the similar steps in expressions (\ref{eq:S20})and (\ref{eq:S21}) we obtain the following two traces,
\begin{eqnarray}
 \sum_{\alpha}[\{\bm{\dot{\iota}}S_{0}({\bm{\rho}}_{in}\times \bm{S}^{\dagger})+
\bm{\dot{\iota}}(\bm{S}\times{\bm{\rho}}_{in})S_{0}^{\dagger}\} ]_{\alpha\alpha}
&=&-\sum_{\{ijk\},\alpha\beta}2\left.\Bigg\{\text{Im}(S_{\alpha\beta,0}S_{\alpha\beta,k}^{*})\rho_{in,\beta,j}  \nonumber \right. \\
&+&\left. \text{Im}(S_{\alpha\beta,0}^{*}S_{\alpha\beta,j})\rho_{in,\beta,k}\right.\Bigg\}\hat{i} \label{eq:S32}
\end{eqnarray}
\begin{eqnarray}
\sum_{\alpha}[\bm{\dot{\iota}}(\bm{S}\rho_{in}\times \bm{S}^{\dagger})]_{\alpha\alpha}
&=&-\sum_{\{ijk\},\alpha\beta}2\text{Im}(S_{\alpha\beta,j}S_{\alpha\beta,k}^{*})\rho_{in,\beta,0}\hat{i} \label{eq:S33}.
\end{eqnarray}
Adding the expression (\ref{eq:S29}-\ref{eq:S33}), we obtain the trace expression (\ref{eq:DP-SD}) of the main text,i.e,
\begin{eqnarray}
\text{tr}[{\bm{\varrho}}_{c}]&=& \sum_{(ijk)_{c},\alpha\beta}2[\text{Re}(S_{0,\alpha\beta}S_{i,\alpha\beta}^{*})-\text{Im}(S_{j,\alpha\beta}S_{k,\alpha\beta}^{*})]\rho_{in,\beta,0}\hat{i} \nonumber \\
&+&\sum_{(ijk)_{c},\alpha\beta}[|S_{0,\alpha\beta}|^{2}+ |S_{i,\alpha\beta}|^{2}- |S_{j,\alpha\beta}|^{2}-|S_{k,\alpha\beta}|^{2}] p_{in,\beta,i}\hat{i} \nonumber \\
&+&\sum_{(ijk)_{c},\alpha\beta}2[\text{Re}(S_{i,\alpha\beta}S_{j,\alpha\beta}^{*})-\text{Im}(S_{0,\alpha\beta}S_{k,\alpha\beta}^{*})]p{in,\beta,j}\hat{i} \nonumber \\
&+&\sum_{(ijk)_{c},\alpha\beta}2[\text{Re}(S_{i,\alpha\beta}S_{k,\alpha\beta}^{*})-\text{Im}(S_{0,\alpha\beta}^{*}S_{j,\alpha\beta})]p_{in,\beta,k}\hat{i}
\end{eqnarray}

In simplifying the various trace expressions obtained in this section of appendix we have used the fact that $\sum_{\alpha\beta}S_{\beta\alpha,i}^{*}S_{\beta\alpha,j}=\sum_{\alpha\beta}S_{\alpha\beta,i}^{*}S_{\alpha\beta,j}$, because $\alpha$ and $\beta$ are dummy indices which are summed over, but one has to be careful and remember that 
$S_{i,\beta\alpha}^{*}\neq S_{i,\alpha\beta}^{*}$.

\subsection{Trace of matrices ${\bm{S}\bm{S}^{\dagger}}$ , ${\bm{S}^{\dagger}\bm{S}}$ and $\text{Tr}[(\bm{S}\bm{S}^{\dagger}-\bm{S}^{\dagger}\bm{S})\bm{\sigma}]$}
\label{sec-S5}
Using the following expressions,
\begin{eqnarray}
 \bm{S}\bm{S}^{\dagger}&=&\left[S_{0}S_{0}^{\dagger}+\bm{S}\cdot\bm{S}^{\dagger}\right]\bm{I}_{2}+\left[(S_{0}\bm{S}^{\dagger}
+\bm{S}S_{0}^{\dagger})+ i(\bm{S}\times \bm{S}^{\dagger}) \right]\cdot \bm{\sigma},
\label{eq:N1} \\
 \bm{S}^{\dagger}\bm{S}&=&\left[S_{0}^{\dagger}S_{0}+\bm{S}^{\dagger}\cdot\bm{S}\right]\bm{I}_{2}+\left[(\bm{S}^{\dagger}\bm{S}_{0}
+S_{0}^{\dagger}\bm{S})+ i(\bm{S}^{\dagger}\times \bm{S}) \right]\cdot \bm{\sigma}.
\label{eq:N2}
\end{eqnarray}
it is straightforward to see that,
\begin{eqnarray} 
\text{Tr}\left[S_{0}S_{0}^{\dagger}+\bm{S}\cdot\bm{S}^{\dagger}\right]\bm{I}_{2}&=&
\text{Tr}\left[S_{0}^{\dagger}S_{0}+\bm{S}^{\dagger}\cdot\bm{S}\right]\bm{I}_{2}\nonumber \\
&=&2 \sum_{i,\alpha, \beta} \left[ (|S_{0,\alpha \beta}|^{2} +|S_{i,\alpha \beta}|^{2})\right] \nonumber \\
&\equiv&\sum_{i,\alpha, \beta} {\cal{C}}_{\alpha\beta}
\label{eq:tr-ss-3} \\ \nonumber \\
\text{Tr}({\bm{S}\bm{S}^{\dagger}\bm{\sigma}})&=&\text{Tr} \left[\{S_{0}\bm{S}^{\dagger}
+\bm{S}S_{0}^{\dagger}\}+ i(\bm{S}\times \bm{S}^{\dagger}) \right] \nonumber \\
&=&\sum_{k,\alpha\beta}[S_{0,\alpha\beta}({S}^{\dagger})_{k,\beta\alpha}+{S}_{k,\alpha\beta}(S^{\dagger})_{0,\beta \alpha}]\hat{k} +\{i\sum_{\alpha\beta,mnk}\epsilon_{mnk}S_{m,\alpha\beta}({S}_{n}^{\dagger})_{\beta\alpha}\} \hat{k} \nonumber \\
%&=&\sum_{k,\alpha\beta}2\text{Re}(S_{0,\alpha\beta}S_{k,\alpha\beta}^{*})\hat{k}+i\sum_{mnk,\alpha\beta}[\epsilon_{mnk}S_{m,\alpha\beta}{S}_{n,\alpha\beta}^{*}] \hat{k} \nonumber \\
&=&\sum_{k,\alpha\beta}2\text{Re}(S_{0,\alpha\beta}S_{k,\alpha\beta}^{*})\hat{k}+i\sum_{\{mnk\},\alpha\beta}[S_{m,\alpha\beta}{S}_{n,\alpha\beta}^{*}-S_{n,\alpha\beta}{S}_{m,\alpha\beta}^{*}]\hat{k} \nonumber \\
&=&\sum_{\{mnk\},\alpha\beta}[2\text{Re}(S_{0,\alpha\beta}S_{k,\alpha\beta}^{*})-2 \text{Im}(S_{m,\alpha\beta}{S}_{n,\alpha\beta}^{*})]\hat{k} \\
&=&\sum_{\{mnk\},\alpha\beta}[2{\cal{Q}}^{(0)}_{mnk,\alpha\beta}]\hat{k}\label {eq:tr-ss-4}
\end{eqnarray}

\appendix

\end{document}